\documentclass[aps,prd,preprint,tightenlines,superscriptaddress,
   showpacs,nofootinbib]{revtex4}
\newcommand{\PRE}[1]{{#1}} 

\usepackage{bm} \usepackage{epsfig}
\usepackage{multirow}

\usepackage{slashed}

\newcommand{\be}{\begin{equation}}
\newcommand{\ee}{\end{equation}}
\newcommand{\beq}{\begin{equation}}
\newcommand{\eeq}{\end{equation}}
\newcommand{\beqa}{\begin{eqnarray}}
\newcommand{\eeqa}{\end{eqnarray}}

\newcommand{\mgut}{m_{\text{GUT}}}
\newcommand{\mplanck}{m_{\text{Planck}}}
\newcommand{\mmess}{M_{\text{mess}}}
\newcommand{\cgrav}{C_{\text{grav}}}

\newcommand{\eqref}[1]{Eq.~(\ref{#1})}

\newcommand{\secref}[1]{Sec.~\ref{sec:#1}}

\newcommand{\appref}[1]{App.~\ref{app:#1}}

\newcommand{\Appref}[1]{Appendix~\ref{app:#1}}

\newcommand{\figref}[1]{Fig.~\ref{fig:#1}}

\newcommand{\tableref}[1]{Table~\ref{table:#1}}

\def\slashchar#1{\setbox0=\hbox{$#1$}           
   \dimen0=\wd0                                 
   \setbox1=\hbox{/} \dimen1=\wd1               
   \ifdim\dimen0>\dimen1                        
      \rlap{\hbox to \dimen0{\hfil/\hfil}}#1
   \else                                        
      \rlap{\hbox to \dimen1{\hfil$#1$\hfil}}/
       \fi}
\def\notslashchar#1{\setbox0=\hbox{$#1$}           
   \dimen0=\wd0                                 
   \setbox1=\hbox{/} \dimen1=\wd1               
   \ifdim\dimen0>\dimen1                        
      \rlap{\hbox to \dimen0{\hfil\phantom{/}\hfil}}#1
   \else                                        
      \rlap{\hbox to \dimen1{\hfil$#1$\hfil}}/
       \fi}

\begin{document}

\preprint{UCI-TR-2009-08}

\title{
\PRE{\vspace*{1.5in}}
{\tt SPICE}: Simulation Package for Including Flavor in Collider Events
\PRE{\vspace*{0.3in}}
}

\author{Guy Engelhard}
\affiliation{Department of Particle Physics, Weizmann Institute
of Science, Rehovot 76100, Israel
\PRE{\vspace*{.2in}}
}

\author{Jonathan L.~Feng}
\affiliation{Department of Physics and Astronomy, University of
California, Irvine, CA 92697, USA
\PRE{\vspace*{.2in}}
}

\author{Iftah Galon}
\affiliation{Physics Department, Technion-Israel Institute of
Technology, Haifa 32000, Israel
\PRE{\vspace*{.5in}}
}

\author{David Sanford}
\affiliation{Department of Physics and Astronomy, University of
California, Irvine, CA 92697, USA
\PRE{\vspace*{.2in}}
}

\author{Felix Yu\PRE{\vspace*{.5in}}}
\affiliation{Department of Physics and Astronomy, University of
California, Irvine, CA 92697, USA
\PRE{\vspace*{.2in}}
}

\date{April 2009}


\begin{abstract}
\PRE{\vspace*{.3in}} We describe {\tt SPICE}: Simulation Package for
Including Flavor in Collider Events.  {\tt SPICE} takes as input two
ingredients: a standard flavor-conserving supersymmetric spectrum and
a set of flavor-violating slepton mass parameters, both of which are
specified at some high ``mediation'' scale.  {\tt SPICE} then combines
these two ingredients to form a flavor-violating model, determines the
resulting low-energy spectrum and branching ratios, and outputs {\tt
HERWIG} and SUSY LesHouches files, which may be used to generate
collider events.  The flavor-conserving model may be any of the
standard supersymmetric models, including minimal supergravity,
minimal gauge-mediated supersymmetry breaking, and anomaly-mediated
supersymmetry breaking supplemented by a universal scalar mass.  The
flavor-violating contributions may be specified in a number of ways,
from specifying charges of fields under horizontal symmetries to
completely specifying all flavor-violating parameters.  {\tt SPICE} is
fully documented and publicly available, and is intended to be a
user-friendly aid in the study of flavor at the Large Hadron Collider
and other future colliders.

\end{abstract}


\maketitle


\section{Program Summary}
\noindent {\it Program title:} {\tt SPICE} \\
{\it Programming language:} {\tt C++} \\
{\it Computer:} Personal computer \\
{\it Operating System:} Tested on Scientific Linux 4.x \\
{\it Keywords:} Supersymmetry, flavor physics, lepton flavor violation \\
{\it PACS:} 11.30.Hv, 12.15.Ff, 14.60.Pq, 12.60.Jv, 13.85.-t \\
{\it External routines:} {\tt SOFTSUSY}~\cite{Allanach:2001kg,Allanach:2009bv} 
and {\tt SUSYHIT}~\cite{Djouadi:2006bz} \\
{\it Nature of problem:} Simulation programs are required to compare
theoretical models in particle physics with present and future data at
particle colliders.  {\tt SPICE} determines the masses and decay branching
ratios of supersymmetric particles in theories with lepton flavor
violation.  The inputs are the parameters of any of several standard
flavor-conserving supersymmetric models, supplemented by flavor-violating
parameters determined, for example, by horizontal flavor symmetries.  The
output are files that may be used for detailed simulation of
supersymmetric events at particle colliders. \\
{\it Solution method:} Simpson's rule integrator, basic algebraic 
computation. \\
{\it Additional comments:} {\tt SPICE} interfaces with {\tt SOFTSUSY} and
{\tt SUSYHIT} to produce the low-energy sparticle spectrum. Flavor mixing
for sleptons and sneutrinos is fully implemented; flavor mixing for
squarks is not included.  \\


\section{Introduction}
\label{sec:intro}

There are two flavor problems in particle physics.  The standard model
(SM) flavor problem is the unexplained pattern of SM fermion masses.
In addition, there is the new physics flavor problem, that is, the
difficulty of solving the gauge hierarchy problem, which inevitably
requires the introduction of new particles at the weak scale, without
generically violating well-known low-energy flavor constraints.
Future colliders, including the Large Hadron Collider (LHC), may shed
light on both the SM and the new physics flavor problems.

Unfortunately, the potential of the LHC in this regard is not well
explored, in part because simulations of new physics typically focus
on the simplest possibilities in which flavor violation is absent.  In
this paper, we describe a new tool, {\tt SPICE}: Simulation Package
for Including Flavor in Collider Events, which is intended as a first
step toward addressing this problem.

In more detail, there are several motivations for developing collider
event simulation tools for models with non-trivial flavor physics:
\begin{itemize}
\item In many new physics frameworks, such as supersymmetry or extra
dimensions, new states are predicted that are governed by the same
flavor or horizontal symmetries as the SM fermions.  Understanding
their masses and mixings may therefore lead to real progress in
identifying the theory of flavor.
\item Although the new physics may be minimally flavor violating, this
is by no means required by current constraints; see, for example,
Refs.~\cite{Feng:2007ke,BarShalom:2007pw,Kribs:2007ac,Nomura:2007ap,%
Nomura:2008pt,BarShalom:2008fq,Nomura:2008gg,Hiller:2008sv}.
As LHC data become available, it is appropriate to relax theoretical
constraints and consider more general frameworks, especially if they
yield unusual signals.
\item The exploration of new physics is often said to consist of three
phases: particle discovery, measurement of particle masses, and
measurement of flavor and other mixings.  In reality, these stages are
unlikely to be completely distinct, and the exploration of flavor
effects may provide new opportunities for particle discovery or
proceed in parallel with precision mass measurements.
\item There are many promising probes of flavor violation at low
energies.  If a positive signal appears in one of them, for example,
$\mu \to e \gamma$, there will be great interest in understanding
whether new weak-scale physics can explain it. In this scenario, a
program like {\tt SPICE} will be helpful to understand what
flavor-violating effects are observed or observable at colliders.
\item Finally, searches for new physics often implicitly assume the
(near) absence of flavor violation. It is important to test their
robustness in the presence of flavor violation.
\end{itemize}

Although {\tt SPICE} is flexible in several ways, in its typical form
{\tt SPICE} reads in parameters for one of the standard
flavor-conserving supersymmetric models and lepton flavor-violating
(LFV) masses specified by the user.  {\tt SPICE} uses the
flavor-conserving model as a spine, then includes the corrections from
LFV terms, determines the new superpartner masses and mass
eigenstates, calculates branching ratios for all relevant
flavor-conserving and flavor-violating decay modes, and outputs the
results in {\tt HERWIG}~\cite{Marchesini:1991ch,Corcella:2000bw,%
Corcella:2002jc,Moretti:2002eu} and SUSY Les Houches Accord (SLHA and
SLHA2)~\cite{Skands:2003cj,Allanach:2008qq} format files, which may be
used to generate collider events.  {\tt SPICE} uses {\tt
SOFTSUSY}~\cite{Allanach:2001kg,Allanach:2009bv} to evolve
supersymmetric parameters through renormalization group equations and
uses {\tt SUSYHIT}~\cite{Djouadi:2006bz} to calculate
flavor-conserving decay widths.

The flavor-conserving model may be any of the standard ones, for
example minimal supergravity (mSUGRA), minimal gauge-mediated
supersymmetry breaking (mGMSB), or anomaly-mediated supersymmetry
breaking supplemented by a universal scalar mass (mAMSB). We review
these and their input parameters below.

The LFV may be specified in a number of ways.  At the ``highest,''
most model-dependent level, a user may specify a horizontal global
symmetry \cite{Froggatt:1978nt,Nir:1993mx,Grossman:1995hk}, the
charges of each of the supermultiplets under this symmetry, and the
Froggatt-Nielsen expansion parameter $\lambda$.  {\tt SPICE} then
populates each LFV mass matrix with entries that have the appropriate
powers of $\lambda$, multiplied by randomly-chosen $\mathcal{O}(1)$
coefficients.  At a slightly lower level, the user may specify each of
the $\mathcal{O}(1)$ coefficients by hand.  And finally, at the lowest
and most model-independent level, a user may specify every LFV term by
hand.

Supersymmetric models with LFV are, of course, just some of the many
possibilities for non-trivial flavor physics at the LHC.  We are led
to consider LFV in supersymmetric models for a variety of reasons.
Supersymmetry is well-motivated by the gauge hierarchy problem and
force unification, both with and without gravity, and there exist
supersymmetric flavor models that can both explain the observed
charged lepton and neutrino masses and mixings and simultaneously
solve the new physics flavor problem.  In addition, and perhaps most
important for the coming data-driven era, supersymmetry is flexible
enough to encompass many diverse experimental signatures, and lepton
flavor violation, as opposed to quark flavor violation, generically
has obvious implications for observable experimental signatures and
event topologies. That said, it would certainly also be interesting to
study hadronic flavor violation, particularly that involving the 3rd
generation, and to generalize these results to non-supersymmetric
frameworks.  A recent study of the observability of squark flavor 
violation at the LHC was done in Ref.~\cite{Kribs:2009zy}.

The program and detailed installation instructions can be found on the
web at the address
\centerline{{\tt http://hep.ps.uci.edu/$\sim$spice} \ .}
In addition, \appref{install} is a step-by-step guide to installing
{\tt SPICE} and gives helpful tips on getting started.

In the following sections, we explain how {\tt SPICE} works.  In
\secref{framework}, we discuss the theoretical framework in more
detail.  In the presence of LFV, many new decay modes open up, but in
most cases, the decay widths are simple generalizations of
flavor-conserving ones.  An exception is the case of
``charge-flipping,'' three-body slepton decays to like-sign leptons
$\tilde{\ell}_i^- \to \tilde{\ell}_j^+ \ell_k^- \ell_m^-$.  We have calculated
these and discussed their interesting phenomenology
elsewhere~\cite{Feng:3body}, but we summarize the main points in
\secref{framework}.

In \secref{prog}, we show how the program works, explaining how to
install the program and specifying the relevant file names and input
and output formats.  To aid first-time users, we also present a simple
example input file and the resulting output.  Our conclusions are
given in \secref{conc}.  Further details regarding our conventions,
Lagrangian terms, decay widths, and program details are given in a
series of appendices.


\section{Theoretical Framework}
\label{sec:framework}

\subsection{Flavor-Conserving Inputs}

The user inputs to {\tt SPICE} consist of a flavor-conserving SUSY
spine supplemented by LFV terms.  The flavor-conserving SUSY spine may
be any model normally available in {\tt SOFTSUSY}
\cite{Allanach:2001kg,Allanach:2009bv}, though in principle any model
could be input with appropriate alterations to the {\tt SOFTSUSY}
code.  The base options and their input parameters are the following:
\begin{itemize}
\item Minimal supergravity (mSUGRA) is specified by
\begin{equation}
\text{mSUGRA: } m_0, M_{1/2}, A_0, \tan\beta, \mgut, \text{sign}(\mu) \ ,
\end{equation}
where $m_0$, $M_{1/2}$, and $A_0$ are the universal scalar, gaugino,
and tri-linear scalar coupling parameters at the grand unified theory
(GUT) scale $\mgut$, $\tan\beta = \langle H_0^u \rangle / \langle
H_0^d \rangle$, and $\text{sign}(\mu)$ is the sign of the Higgsino
mass parameter $\mu$.

\item Minimal gauge-mediated supersymmetry breaking (mGMSB) is
  specified by
\begin{equation}
\text{mGMSB: } N_5, \mmess, \Lambda, \cgrav, \tan\beta,
\text{sign}(\mu) \ ,
\end{equation}
where $N_5$ is the number of ${\bf 5} + {\bf \overline{5}}$ multiplets
in the messenger sector, $\mmess$ is the messenger mass scale,
$\Lambda = F_S/\langle S \rangle$ sets the mass scale for the SM
superpartners, $\cgrav = m_{\tilde{G}} / (F/\sqrt{3} \mplanck)$ is the
gravitino mass in units of its mass in the minimal case in which there
is only one SUSY-breaking $F$ term, and $\tan\beta$ and
$\text{sign}(\mu)$ are as above.

\item Minimal anomaly-mediated supersymmetry breaking (mAMSB) is
  specified by
\begin{equation}
\text{mAMSB: } m_0, m_{3/2}, \tan\beta, \mgut, \text{sign}(\mu) \ ,
\end{equation}
where $m_0$ is the universal scalar mass, motivated by the tachyonic
slepton problem and added to all scalars, including the Higgs scalars,
$m_{3/2}$ is the gravitino mass, which sets the scale for all SM
superpartners, and $\tan\beta$, $\mgut$, and $\text{sign}(\mu)$ are as
above.

\end{itemize}

\subsection{Flavor-Violating Inputs: Model-Independent Approach}

The LFV terms are specified by a parameter $x$ and three $3\times 3$
matrices $m_E$, $X_L$, and $X_R$.  All of these matrices are assumed
real, and $X_L$ and $X_R$ are necessarily symmetric.  In the most
model-independent formulation accommodated by {\tt SPICE}, LFV is
therefore completely specified by 22 numbers: $x$, 9 for $m_E$, 6 for
$X_L$, and 6 for $X_R$.

The matrix $m_E$ is the SM charged lepton mass matrix.  $X_L$ and
$X_R$ determine the slepton mass matrices through
\begin{eqnarray}
M_{\tilde{\nu}}^2 &=& m_{\tilde{L}}^2 {\mathbf 1} + x \tilde{m}^2 X_L
\\
M_{\tilde{E}_L}^2 &=& m_{\tilde{L}}^2 {\mathbf 1} + m_E m_E^{\dagger}
+ x \tilde{m}^2 X_L \\ 
M_{\tilde{E}_R}^2 &=& m_{\tilde{R}}^2 {\mathbf 1} + m_E^{\dagger} m_E
+ x \tilde{m}^2 X_R \ ,
\end{eqnarray}
where $m_{\tilde{L}}^2$ and $m_{\tilde{R}}^2$ are the
flavor-conserving contributions to the left- and right-handed
sleptons, and $\tilde{m}^2$ characterizes the size of
flavor-conserving contributions to the slepton masses.  For
concreteness we take $\tilde{m}^2$ to be equal to the average of the
diagonal elements of $m_{\tilde{L}}^2$ \footnote{For a
flavor-conserving spine where the slepton mass-squared matrices are
not proportional to the identity, $\tilde{m}^2$ is thus equal to the
average left slepton mass-squared.}.  The parameter $x$
specifies the size of the LFV effects relative to the
flavor-conserving parameters.

Several simplifying assumptions are encoded in this formulation of the
LFV effects:
\begin{enumerate}
\item In assuming $m_E$, $X_L$, and $X_R$ are real, we do not include
  CP violation from these matrices.  
\item The flavor-violating contributions to left- and right-handed
  sleptons are of the same order, as set by $x \tilde{m}^2$.  This is
  what one would expect of gravitational contributions, which are
  chirality-blind.
\item In general, there should also be LFV $A$-terms.  We assume these
  are negligible at the mass scale where these SUSY-breaking masses
  are generated.  Note, however, that $A$-terms are generated through
  RG evolution, and such effects are included.
\item In general, there should also be a non-trivial LFV neutrino mass
  matrix.  However, since we are primarily interested in colliders,
  where neutrino flavor is unobservable, we simply assume that the
  neutrino masses vanish, and so the neutrino mass and gauge
  eigenstates are identical.  Note that sneutrinos may have observable
  mixings, and these are included.
\end{enumerate}

\subsection{Flavor-Violating Inputs: Flavor Symmetry Approach}

An attractive possibility is that the LFV terms are determined by
horizontal symmetries.  In this approach, the 3 LFV matrices are
\begin{eqnarray}
m_E &=& m_{\ell} \left( \begin{array}{ccc}
c_1 \lambda^{n_1} & c_2 \lambda^{n_2} & c_3 \lambda^{n_3} \\
c_4 \lambda^{n_4} & c_5 \lambda^{n_5} & c_6 \lambda^{n_6} \\
c_7 \lambda^{n_7} & c_8 \lambda^{n_8} & c_9 \lambda^{n_9}
\end{array} \right) \\
X_L &=& \left( \begin{array}{ccc}
c_{10} \lambda^{n_{10}} & c_{11} \lambda^{n_{11}} & c_{12} \lambda^{n_{12}} \\
c_{11} \lambda^{n_{11}} & c_{13} \lambda^{n_{13}} & c_{14} \lambda^{n_{14}} \\
c_{12} \lambda^{n_{12}} & c_{14} \lambda^{n_{14}} & c_{15} \lambda^{n_{15}}
\end{array} \right) \\
X_R &=& \left( \begin{array}{ccc}
c_{16} \lambda^{n_{16}} & c_{17} \lambda^{n_{17}} & c_{18} \lambda^{n_{18}} \\
c_{17} \lambda^{n_{17}} & c_{19} \lambda^{n_{19}} & c_{20} \lambda^{n_{20}} \\
c_{18} \lambda^{n_{18}} & c_{20} \lambda^{n_{20}} & c_{21} \lambda^{n_{21}}
\end{array} \right) ,
\end{eqnarray}
where $m_{\ell}$ is the lepton mass scale, the $c_i$ are ${\cal O}(1)$
coefficients, $\lambda$ is the Froggatt-Nielsen expansion parameter,
and the exponents $n_i$ are determined by supermultiplet charges.

{\tt SPICE} accommodates this possibility by providing an alternative
way to specify the LFV parameters in the case of $U(1)^n$ flavor
models, with breaking parameters of the same size.  For example, for
$U(1) \times U(1)$, one sets $N_{\text{charges}} = 2$, and specifies
the two U(1) charges for the six multiplets $L_1$, $L_2$, $L_3$,
$E_1$, $E_2$, and $E_3$.  {\tt SPICE} then determines the correct
exponent $n_i$ for each entry of the three matrices.  {\tt SPICE} may
also determine the coefficients $c_i$ randomly or these may be set by
the user.

\subsection{Spectrum and Decay Calculations}

Once the flavor-conserving and flavor-violating parameters are input,
the model is completely specified.  Formally, one should then evolve
the mass parameters from the scale they are generated, adding in new
contributions at the appropriate scale, until one reaches the weak
scale.  In {\tt SPICE}, we instead add the LFV contributions to the
flavor-conserving contributions at the scale at which the
flavor-conserving contributions are generated, and then RG evolve the
result to the weak scale.  In the case that the flavor-conserving and
LFV contributions are generated at the same scale, as in the case of
mSUGRA or mAMSB spines with LFV generated at $\mgut$, this is the
correct prescription.  If the generation scales differ, for example,
as in the case of a mGMSB spine with gravity-mediated LFV effects, our
prescription is not formally correct.  Even in this case, however, we
expect that the dominant difference resulting from this approximation
may be absorbed into the LFV ${\cal O}(1)$ parameters, which are not
completely specified anyway.

Given the slepton masses and mixings determined above, we then
calculate all branching ratios involving these LFV effects.  Most of
the flavor-violating decay widths are, at least to tree level, fairly
simple generalizations of the flavor-conserving widths.  New effects
appear in the 3-body decays of sleptons both to like-sign leptons,
$\tilde{\ell}_i^- \to \tilde{\ell}_j^+ \ell_k^- \ell_m^-$, and to
opposite-sign leptons, $\tilde{\ell}_i^- \to \tilde{\ell}_j^- \ell_k^-
\ell_m^+$.  Previous work on these
decays~\cite{Ambrosanio:1997bq,Kraml:2007sx} has assumed flavor
conservation. In the presence of LFV, however, the lepton pair may be
identical particles or charge conjugates of the same generation,
leading to additional interference terms.  We have calculated these
decay widths in the presence of LFV and arbitrary left-right slepton
mixing in another work~\cite{Feng:3body} and included these in {\tt
SPICE}.

Currently {\tt SPICE} discards decays to gravitinos.  Such decays are
typically either kinematically inaccessible or too slow to be relevant
at colliders.  Even in the cases where the model used for the spine
predicts fast decays to gravitinos, as in the case of mGMSB with a low
supersymmetry-breaking scale, the existence of significant LFV
contributions implies that in the full LFV theory, supersymmetry is
broken at a high scale, and decays to gravitinos are negligible.


\section{Program}
\label{sec:prog}

\subsection{Procedure}
\label{subsec:proc}
\begin{figure}
\includegraphics[scale=0.5]{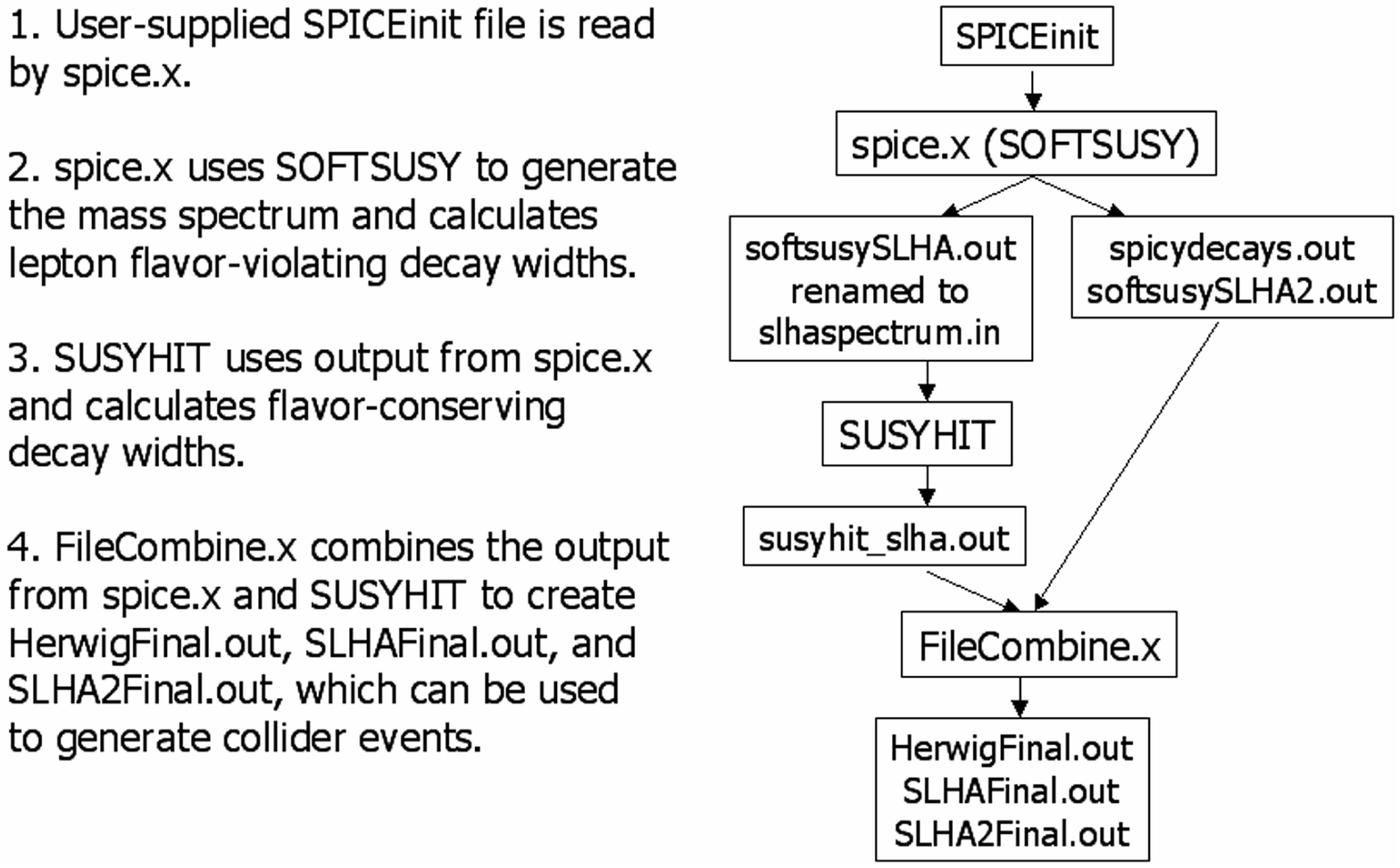}
\caption{Schematic flow chart for {\tt SPICE}, showing the progression
from the {\tt SPICEinit} input file to the final output files {\tt
HerwigFinal.out}, {\tt SLHAFinal.out}, and {\tt SLHA2Final.out}.  The
explicit interfacing with {\tt
SOFTSUSY}~\cite{Allanach:2001kg,Allanach:2009bv} and {\tt
SUSYHIT}~\cite{Djouadi:2006bz} is also shown.}
\label{fig:flowchart}
\end{figure}

The program is designed to take in one input file which specifies all
of the input parameters.  This file, by default, is called {\tt
SPICEinit} and is located in the {\tt spice/} directory.  The default
input file is identical to the example input file presented below.
These input parameters are used by {\tt SOFTSUSY}
\cite{Allanach:2001kg,Allanach:2009bv} to generate a particle mass
spectrum, which is passed to {\tt SUSYHIT}~\cite{Djouadi:2006bz} to
create a SUSY Les Houches Accord (SLHA) file with the traditional
flavor-conserving decay table.  Decay widths of the flavor-generalized
decay channels involving sleptons and sneutrinos are calculated
separately within the {\tt SOFTSUSY} executable, which generates an
intermediate file containing the flavor-generalized mass spectrum and
relevant decays.  The {\tt FileCombine} subprogram merges the {\tt
SUSYHIT} output with this intermediate file to generate the file {\tt
HerwigFinal.out}, which is an input file for the {\tt Herwig} event
generator
\cite{Marchesini:1991ch,Corcella:2000bw,Corcella:2002jc,Moretti:2002eu}
and both {\tt SLHAFinal.out} and {\tt SLHA2Final.out}, which are SLHA
\cite{Skands:2003cj} and SLHA2~\cite{Allanach:2008qq} formatted files
appropriate for input to Monte Carlo event generators.  The program
flow is diagrammed schematically in \figref{flowchart}.

The input file for the program contains two sets of input parameters
--- the flavor-conserving SUSY spine, and the flavor-violating
effects.  The first set specifies the SUSY breaking scenario
parameters according to the {\tt SOFTSUSY}
\cite{Allanach:2001kg,Allanach:2009bv} specifications, which are
briefly reviewed below.  The second set describe high energy flavor
mixing boundary conditions.  The default input file is presented below
as an example\footnote{This example file uses the U(1) horizontal
charges of Model B in Ref.~\cite{Feng:2007ke}.}.

\subsubsection*{Example Input File}
\label{subsubsec:examplein}
\noindent {\tt
\indent gmsb 4 2.0e6 5.0e4 1.0 10 1 \\
\indent x 0.1 \\
\indent lambda 0.2 \\
\indent nCharges 2 \\
\indent L1 2 0 \\
\indent L2 0 2 \\
\indent L3 0 2 \\
\indent E1 2 1 \\
\indent E2 2 -1 \\
\indent E3 0 -1 \\
\indent Lep -0.13854 2.24310 -0.4399 1.84877 1.53904 -0.56145 0.89942 0.53290 1.45314 \\
\indent XL 0.98914 -1.2001 1.40588 -1.2001 -0.470473 2.67457 1.40588 2.67457 -0.22933 \\
\indent XR -0.5889 2.8415 -0.2321 2.8415 -0.43167 1.3267 -0.2321 1.3267 0.55438 \\
}
\begin{itemize}
\item{\bf Flavor conserving parameters:} \newline
\noindent We incorporate the three SUSY breaking scenarios implemented
by {\tt SOFTSUSY} as options for the flavor-conserving SUSY spine.
These are mSUGRA, mGMSB, and mAMSB, as described in
\secref{framework}.  The first line of the script file specifies which
spine is used with an identifier of ``sugra'', ``gmsb'', or ``amsb'',
and then parameters needed for that given scenario.  The ordering is
consistent with that used in \secref{framework} and the {\tt SOFTSUSY}
manual~\cite{Allanach:2001kg,Allanach:2009bv}.  Note that the text
labels for the SUSY breaking scenario is case sensitive.  These are
summarized as follows: \newline
\hspace*{0.5in}{\tt sugra <m0> <m12> <a0> <tanb> <mgut> <sgnMu>}
\newline
\hspace*{0.5in}{\tt gmsb <n5> <mMess> <lambda> <cgrav> <tanb> <sgnMu>}
\newline
\hspace*{0.5in}{\tt amsb <m0> <m32> <tanb> <mgut> <sgnMu>} \newline

\item{\bf Flavor mixing parameters} \newline
\noindent The remainder of the input file is used to describe the
slepton flavor mixing in the model in the context of U(1) horizontal
charges as described in Ref.~\cite{Feng:2007ke} and reviewed in
\secref{framework}.  The first two parameters are $x$ and $\lambda$,
and the next integer nCharges specifies the number of U(1) charges for
the leptons.  Note that text labels are required as placeholders
before each variable, although the actual label text is arbitrary for
all entries except the spine identifier and the ``Lep'' identifier
which may be changed to ``random'' to generate random coefficients
(see below).  After the number of charges is passed, the charge
assignments for each generation of left-handed sleptons and
right-handed sleptons must be given.  Lastly, the user can specify the
$\mathcal{O}(1)$ coefficients to be used in the lepton, left-handed
slepton, and right-handed slepton mass matrices.  The blocks ``Lep'',
``SLepXL'', and ``SLepXR'' define the nine $\mathcal{O}(1)$
coefficients for each $3 \times 3$ mass matrix, according to the
pattern
\begin{equation}
\text{Lep}\ \ c_1\ \ c_2\ \ c_3\ \ c_4\ \ c_5\ \ c_6\ \
c_7\ \ c_8\ \ c_9 \ \to
\left( \begin{array}{ccc}
c_1 \lambda^{n_1} & c_2 \lambda^{n_2} & c_3 \lambda^{n_3} \\
c_4 \lambda^{n_4} & c_5 \lambda^{n_5} & c_6 \lambda^{n_6} \\
c_7 \lambda^{n_7} & c_8 \lambda^{n_8} & c_9 \lambda^{n_9}
\end{array} \right).
\end{equation}
``SLepXL'' and ``SLepXR'' are symmetric matrices and must be defined
with symmetric coefficients, while the ``Lep'' matrix has arbitrary
coefficients.  Alternatively, the randomly generated coefficients may
be generated by replacing the last three lines with \newline
\hspace*{0.5in}{\tt random <rseed> <sigma>} \newline
where {\tt <rseed>} is an integer, used as a random seed to generate
$\mathcal{O}(1)$ numbers of the form $\pm \exp (a)$, where $a$ has a
gaussian distribution with a mean of 0 and a $\sigma$ of {\tt
<sigma>}.  These coefficients have a 50\% probability of being
negative.
\end{itemize}
Besides the default example file, there are several example input
files provided in the sub-directory {\tt spice/Examples/} that
demonstrate various charge assignments following the models A-D
prescription in Ref.~\cite{Feng:2007ke}.  To use these, rename them to
the default input file name {\tt SPICEinit} in the {\tt spice/}
directory, or modify the {\tt CalcDecays} command script
appropriately.

\subsection{{\tt SOFTSUSY} and {\tt SUSYHIT} Details}
\label{subsec:details}
{\tt SPICE} relies upon {\tt SOFTSUSY}
\cite{Allanach:2001kg,Allanach:2009bv}, using {\tt SOFTSUSY} for
renormalization with flavor-general boundary conditions.  The lepton
flavor mixing is rotated into slepton and sneutrino mass matrices at
the high scale before renormalization is performed; the rotation of
the neutrino masses is neglected along with the neutrino masses
themselves.  In general, diagonalizing the lepton mass matrix requires
rotating the trilinear slepton couplings as well, but this is
unnecessary in our case as we assume negligible trilinear
contributions at the high scale.  Using the flavor general boundary
conditions in {\tt spice.cpp}, {\tt SOFTSUSY} iteratively solves the
renormalization group equations (RGEs) and outputs a SUSY mass
hierarchy in SLHA and SLHA2 format to {\tt softsusySLHA.out} and {\tt
softsusySLHA2.out}, respectively; the SLHA file is passed to {\tt
SUSYHIT} as {\tt slhaspectrum.in}.  We do not force gauge coupling
unification, the GUT scale is set at $m_{\text{GUT}} = 2 \times
10^{16}$ GeV initially, we use 2-loop Higgs formulas for physical
masses, and we generally follow {\tt SOFTSUSY} program defaults and
conventions as much as possible.  For more details on the workings of
{\tt SOFTSUSY}, refer to the {\tt SOFTSUSY} manual
\cite{Allanach:2001kg,Allanach:2009bv}.

{\tt SUSYHIT} takes {\tt slhaspectrum.in} and calculates non-flavor
violating decays of the entire SUSY spectrum, according to its
subroutines {\tt SDECAY} and {\tt HDECAY}.  Details can be found in
the {\tt SUSYHIT} reference manul~\cite{Djouadi:2006bz}.  The output
file appends the calculated decay table to the input file {\tt
slhaspectrum.in}, and this {\tt SUSYHIT} output file is passed to {\tt
FileCombine}.

\subsection{Decay Calculation Files}
\label{subsec:progfiles}
The calculations of new flavor violating decays is performed within
the {\tt SPICE} executable with the code contained in {\tt
  DecayFormat.cpp}, {\tt DecayCalc.cpp} and {\tt
  ThreeBodyIntegrals.cpp} and their respective {\tt .h} header files.
The content of these files is:

\begin{itemize}
\item The {\tt DecayFormat} class:
\begin{itemize}
\item Output of particle mass and lifetime data
\item Output of a standard format of two- and three-body decays
\item Output of SUSY parameters in {\tt HERWIG} format
\item A public method which chains these methods together to produce a
  pseudo-{\tt HERWIG} file\footnote{This pseudo-{\tt HERWIG} file
  differs from a true {\tt HERWIG} file since it lacks SUSY decays
  beyond those involving sleptons and sneutrinos, it does not have the
  appropriate whitespace formatting, and it uses Particle Data Group
  (PDG) particle identification codes instead of {\tt HERWIG} particle
  codes.}
\end{itemize}
\item The {\tt DecayCalc} class:
\begin{itemize}
\item Retrieval of the slepton/sneutrino mass matrices from {\tt
  FlavourMssmSoftsusy}
\item Calculations for all decays involving sleptons and sneutrinos
\item Functions for Lagrangian coefficients and three-body decay
  integral multiplicative factors
\item Loop over all possible sleptons/leptons for the given decays and
  output via {\tt DecayFormat} class
\end{itemize}
\item The {\tt ThreeBodyIntegrals} class:
\begin{itemize}
\item Functions for the integrands of the three body decay integrals
\item A Simpson's rule integration method
\item Methods to calculate the normal and interference term integrals
  given a vector containing multiplicative factors
\end{itemize}
\end{itemize}

The decay calculations (20 in total) are numbered according to
Table~II in \appref{decaytable}.  These replace all the
standard decays calculated by {\tt SUSYHIT} that involve
sleptons/sneutrinos as either decaying particle or product.  The
routine is called from {\tt spice.cpp} using the {\tt
OutputHerwigFile} method of a {\tt DecayFormat} object, which in turn
creates a {\tt DecayCalc} object and proceeds to call the other global
output functions to generate a file in pseudo-{\tt HERWIG} format
containing the slepton/sneutrino decays.  This method also generates
the mass spectrum and model parameters for inclusion in the final
output file appropriate for Monte Carlo event generators.  There are
also several decay modes listed in Table~III in
\appref{decaytable} that are irrelevant or not viable in the models we
have studied but may be viable in other scenarios; instructions on how
to insert new decay modes are presented in \appref{program}.

\subsection{Merging Flavor Violation and SUSYHIT Files}
\label{subsec:merging}
The final component of {\tt SPICE} is the {\tt FileCombine}
subroutine, which takes the flavor-violating slepton/sneutrino decays
file and merges it with the other decays calculated by {\tt SUSYHIT}.
{\tt FileCombine} takes the particle mass info and model parameters
from the {\tt softsusySLHA.out} file, along with the slepton-sneutrino
decay table.  It then reads through the {\tt SUSYHIT} decay file and
stores all decays that do not involve sleptons or sneutrinos.

\subsection{Output}
\label{subsec:output}
The program outputs the stored data as a standard {\tt HERWIG} input
file
\cite{Marchesini:1991ch,Corcella:2000bw,Corcella:2002jc,Moretti:2002eu}
and both SLHA- and SLHA2-formatted files for input to other Monte
Carlo event simulation packages.  The first portion of the {\tt
HERWIG} file contains particle codes, masses, and lifetimes; the bulk
contains branching fractions for particle decays, and the final
section contains model parameters\footnote{The SLHA files are
structured similarly; for details, see
Refs.~\cite{Skands:2003cj,Allanach:2008qq}.}.  It contains all
feasible (and large) flavor-violating slepton and sneutrino decays.
\begin{center}
\begin{figure}[tb]
\includegraphics[scale=0.55]{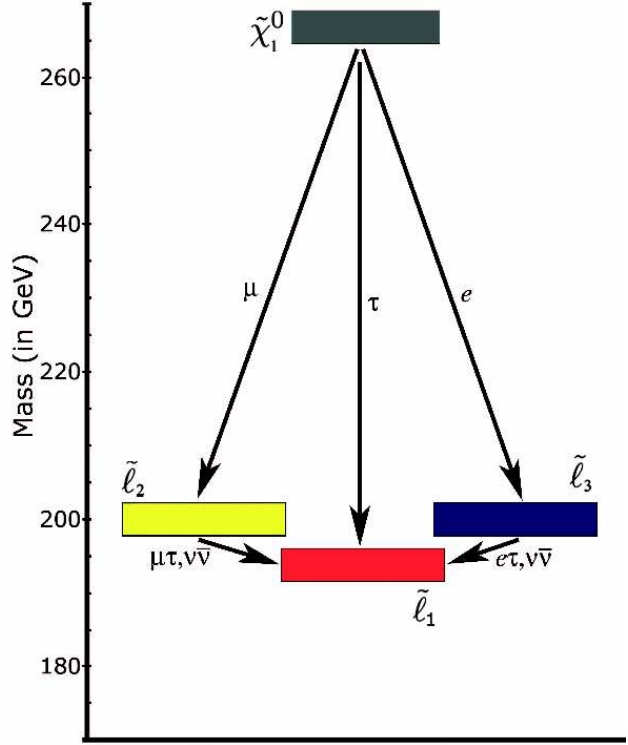}
\caption{Low-energy spectrum and decay modes for the flavor-conserving
GMSB spine specified in the {\tt SPICEinit} example input file.
Flavor-violating effects were not included, and decays to gravitino
are ignored.  For the slepton shadings (colors), stau flavor is medium
gray (red), smuon flavor is light gray (yellow), and selectron flavor
is dark gray (blue).  SUSY decays proceed along arrows and are labeled
by the associated SM daughter particles.  The relevant masses are
$\tilde{\chi}_1^0$ = 266.7 GeV, $\tilde{\ell}_3^-$ = 200.1 GeV,
$\tilde{\ell}_2^-$ = 200.1 GeV, and $\tilde{\ell}_1^-$ = 193.6 GeV.}
\label{fig:lowendFC}
\end{figure}
\end{center}
\begin{center}
\begin{figure}[tb]
\includegraphics[scale=0.55]{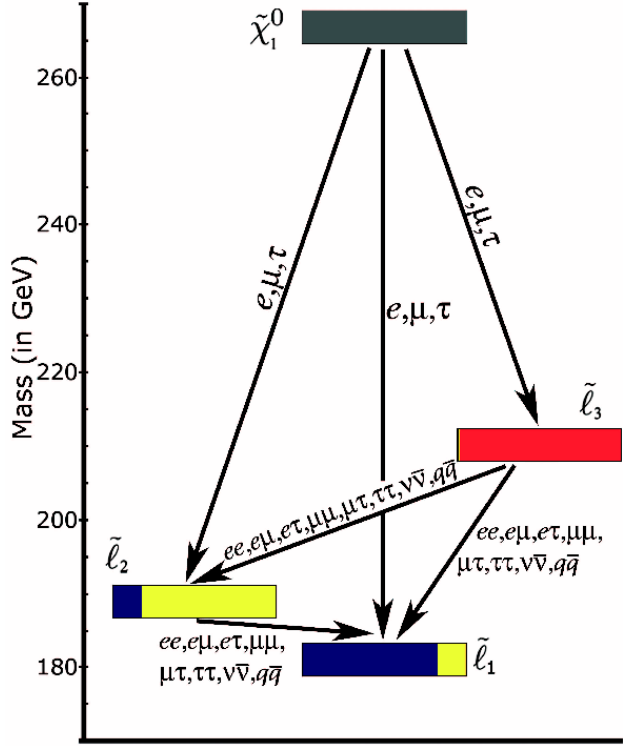}
\caption{Low-energy spectrum and decay modes for the full
flavor-violating model specified in the {\tt SPICEinit} example input
file.  Decays to gravitino are ignored.  Flavor-violating effects are
included and serve to modify slepton masses, mix the gauge
eigenstates, and introduce many new decay modes.  For the slepton
shadings (colors), stau flavor is medium gray (red), smuon flavor is
light gray (yellow), and selectron flavor is dark gray (blue).  SUSY
decays proceed along arrows and are labeled by the associated SM
daughter particles.  The relevant masses are $\tilde{\chi}_1^0$ =
266.7 GeV, $\tilde{\ell}_3^-$ = 210.0 GeV, $\tilde{\ell}_2^-$ = 188.7
GeV, and $\tilde{\ell}_1^-$ = 179.8 GeV, and the branching ratios are
listed in \tableref{FVtable}.}
\label{fig:lowendFV}
\end{figure}
\end{center}

\begin{center}
\begin{table}[b]
\begin{tabular}{|c|c || l|l|l|l|l|l|l|l|l|l|}
\hline
  & {\bf Mass} & {\bf Mode} & $\tilde{\ell}_1^\pm e^\mp$ 
& $\tilde{\ell}_2^\pm \mu^\mp$ & $\tilde{\ell}_3^\pm \tau^\mp$ 
& $\tilde{\ell}_1^\pm \mu^\mp$ & $\tilde{\ell}_2^\pm e^\mp$ 
& & 
& & 4 more \\
$\tilde{\chi}_1^0$ & 266.7 & {\bf B.R.} & 35.4\% 
& 29.5\% & 20.6\% 
& 7.7\% & 6.5\% 
& &
& & 0.3\% \\
\hline
            &            &                  & $\tilde{\ell}_1^+ e^- \tau^-$
& $\tilde{\ell}_1^- \tau^- e^+$ & $\tilde{\ell}_1^+ \mu^- \tau^-$ 
& $\tilde{\ell}_2^+ \mu^- \tau^-$ & $\tilde{\ell}_1^- \tau^- \mu^+$
& $\tilde{\ell}_2^- \tau^- \mu^+$ & $\tilde{\ell}_2^+ e^- \tau^-$ 
& $\tilde{\ell}_2^- \tau^- e^+$ & 50 more \\
$\tilde{\ell}_3^-$ & 210.0 &                       & 44.3\%
& 23.9\% & 9.5\%
& 8.4\% & 5.2\%
& 4.8\% & 1.9\%
& 1.1\% & 0.9\% \\
\hline
            &            &                  & $\tilde{\ell}_1^+ e^- \mu^-$
& $\tilde{\ell}_1^- \mu^- e^+$ & $\tilde{\ell}_1^+ e^- e^-$ 
& $\tilde{\ell}_1^+ \mu^- \mu^-$ & $\tilde{\ell}_1^- e^- e^+$
& $\tilde{\ell}_1^- \mu^- \mu^+$ & $\tilde{\ell}_1^- e^- \mu^+$
& & 22 more \\
$\tilde{\ell}_2^-$ & 188.7 &                       & 27.8\%
& 21.6\% & 20.0\%
& 19.7\% & 4.7\%
& 4.7\% & 1.0\%
& & 0.5\% \\
\hline
\end{tabular} \newline
\caption{List of masses (in GeV) and branching ratios for decaying
particles $\tilde{\chi}_1^0$, $\tilde{\ell}_3^-$, and
$\tilde{\ell}_2^-$ using the {\tt SPICEinit} example input file.  Only
branching ratios larger than 1.0\% are individually listed, and decays
to gravitino are ignored.  This table refers to the spectrum depicted
in \figref{lowendFV}.}
\label{table:FVtable}
\end{table}
\end{center}
For the example input file in Subsection \ref{subsec:proc}, we provide
some comments about the expected output and how our example model
differs from traditional flavor-conserving models.  Using only the
flavor-conserving spine of the example model, for instance, we derive
the traditional low-energy spectrum and straightforward decay channels
depicted in \figref{lowendFC}.  When {\tt SPICE} calculates the full
flavor-violating example model, however, the decays modes acquire new
complexity and richness: these modes are diagrammed in
\figref{lowendFV} and listed in \tableref{FVtable}.  We note that the
flavor mixing parameters for the $X_R$ matrix in the example input
file indicate that the two lightest sleptons are predominantly
selectron and smuon, which is confirmed by the branching ratios for
$\tilde{\chi}_1^0 \to \tilde{\ell}_2^\pm \ell^\mp, \tilde{\ell}_1^\pm
\ell^\mp$.  Thus, while traditional expectations favor the lightest
slepton to be predominantly stau, given the large Yukawa coupling of
the third generation, the lightest sleptons can instead generically be
combinations of selectron and smuon in flavor-violating scenarios.
More importantly, the wealth of new information in flavor-violating
two- and three-body modes indicates it may be possible to measure
flavor couplings from early collider data, if positive signals for
SUSY are discovered.

\section{Conclusion}
\label{sec:conc}

We have presented the program {\tt SPICE}: Simulation Package for
Including Flavor in Collider Events intended for use in studies of
lepton flavor violation (LFV).  {\tt SPICE} simulates hybrid models
where flavor-violating gravitational effects are added to a
flavor-conserving SUSY spine.  The program contains preset options for
generating lepton flavor violation based on U(1) horizontal
symmetries, but it also allows for input of more general
flavor-violating effects.  It interfaces with {\tt SOFTSUSY}, which
generates the low-energy mass hierarchy, and {\tt SUSYHIT}, which
calculates decays unconnected to LFV.  {\tt SPICE} automatically
calculates all kinematically allowed decays involving sleptons and
sneutrinos with LFV, and also calculates three-body slepton decays,
which are important in models with a long-lived slepton.  Given an
input model with LFV, {\tt SPICE} outputs the mass hierarchy and decay
table with LFV effects, which can be used to generate collider events
using Monte Carlo simulation packages.  {\tt SPICE} is freely
available under the Gnu Public License, and can be downloaded at {\tt
  http://hep.ps.uci.edu/$\sim$spice} .

There are several possible extensions to {\tt SPICE}.  Foremost among
these is the inclusion of quark sector flavor violation, which, while
not expected to impact collider signals as obviously as some possible
LFV signals, also provides interesting effects.  Another valuable
extension would be the inclusion of low-energy effects, including
neutrino masses and mixings and comparison to low-energy constraints
such as $\mu \to e \gamma$.  Another possibility is the flavor
generalization and inclusion of three-body decays involving sneutrinos
as detailed in Ref.~\cite{Kraml:2007sx}.  In the case of a long-lived
slepton, there is also the possibility of a decay of the form
$\tilde{\ell}_i \to \tilde{\ell}_j \gamma$ generated through LFV
loops, which could be considered.

\section*{Acknowledgements}

We are grateful to Y.~Nir and Y.~Shadmi for many helpful discussions
and careful readings of the manuscript.  We would also like to thank
S.~French and C.~Lester for helpful advice and comments about program
applications.  The work of JLF, DS, and FY was supported in part by
NSF grants PHY--0239817 and PHY--0653656 and the Alfred P.~Sloan
Foundation. IG thanks the UC Irvine particle theory group for their
hospitality while this work was in progress.  This research was
supported in part by the United States-Israel Binational Science
Foundation (BSF) under grant No.~2006071.  The research of IG was also
supported by the Israel Science Foundation (ISF) under grant
No.~1155/07.


\appendix
\section*{Appendices}

This section of the paper provides all details relevant to running
{\tt SPICE}.  \Appref{install} outlines the installation procedure and
provides general tips on getting started, as well as basic
troubleshooting solutions.  Detailed command-line instructions can be
found in the {\tt README} file on the website.  \Appref{conventions}
covers the notation and conventions used in {\tt SPICE}.
\Appref{lagrangians} defines the relevant portions of the Lagrangian
for flavor-general sleptons and defines Lagrangian coefficients to
simplify the forms of the decay modes.  \Appref{decaytable} enumerates
the flavor-generalized decays used and calculated by {\tt SPICE}, as
well as the decays {\tt SPICE} ignores.  \Appref{twobody} gives
formulas for flavor-generalized two-body decays.  \Appref{threebody}
discusses the three-body decays; the full formulas are presented in
Ref.~\cite{Feng:3body}.  \Appref{program}, finally, discusses the
structure of the program and details about the integration between
{\tt SPICE}, {\tt SOFTSUSY}, and {\tt SUSYHIT}.

\section{Installation Instructions and Troubleshooting}
\label{app:install}

\subsection{Installing and Running {\tt SPICE}}

{\tt SPICE} can be downloaded from {\tt
http://hep.ps.uci.edu/$\sim$spice} .  {\tt SPICE} requires both {\tt
SOFTSUSY} ({\tt http://projects.hepforge.org/softsusy}) and {\tt
SUSYHIT} ({\tt http://lappweb.in2p3.fr/$\sim$muehlleitner/SUSY-HIT})
to be installed in order to function properly.  After obtaining and
unzipping all three packages, use the {\tt SPICE} {\tt Makefile} in
the {\tt spice/} directory to build the executables {\tt spice.x} and
{\tt FileCombine.x}, and also run the default {\tt SUSYHIT} {\tt
Makefile} to compile {\tt SUSYHIT}\footnote{Please note that {\tt
spice.x} contains the relevant portions of {\tt SOFTSUSY}, so a
separate {\tt SOFTSUSY} build is not required.}.  The command script
{\tt CalcDecays}, located in the {\tt spice/} directory, handles
program execution and movement of input and output files: by default,
it uses the file {\tt SPICEinit} for input parameters.  The output of
{\tt SPICE} is a Herwig input file {\tt HerwigFinal.out}, a SUSY
LesHouches Accord file {\tt SLHAFinal.out}, and a SLHA2 file {\tt
SLHA2Final.out}, all located in {\tt spice/FileCombine/}.  When
running the default {\tt SPICEinit} file, the program output should
match the example output files in the {\tt spice/FileCombine/}
directory.  Additional input files, which demonstrate the versatility
of {\tt SPICE}, can be found in {\tt spice/Examples/}.

\subsection{Troubleshooting}

By default, {\tt SPICE} assumes a particular directory structure, with
the {\tt spice/}, {\tt softsusy/}, and {\tt susyhit/} directories all
located in the same parent directory, {\it i.e.}, {\tt
Desktop/spice/}, {\tt Desktop/softsusy/}, and {\tt
Desktop/susyhit/}.  Any other directory structure or different
directory names can be accomodated by making the appropriate changes
to the directory path names in the {\tt SPICE} {\tt Makefile} and
command script {\tt CalcDecays}.

{\tt SPICE} was tested with the standard {\tt SOFTSUSY} 3.0.2. and
{\tt SUSYHIT} 1.3 builds, so a customized build of either program's
core files may cause errors.  In addition, {\tt SPICE} was compiled
using the {\tt g++-3.4} compiler, so use of a different compiler may
require changes to the {\tt SPICE} {\tt Makefile}.  In particular, for
the {\tt gcc-4.2} compiler, the user should remove the {\tt -ff90}
option in the {\tt Makefile} and replace the {\tt -lg2c} inline tag
with {\tt -lgfortran}.  Further installation details are provided in
the {\tt README} file included with {\tt SPICE}.

\section{Conventions}
\label{app:conventions}

In this section we use the spacetime metric ($+$$-$$-$$-$) and
fermionic propagators of the form $\left( \slashed{p} - m \right) /
\left(p^2 - m^2 \right)$.  Our projection operators are $P_{L,R} = (1
\mp \gamma_5)/2$.  The constants $g$ and $g'$ are the standard weak
scale SU(2) and U(1) gauge couplings; we break from {\tt SOFTSUSY}
convention and use the typical standard model value for $g'$ rather
than the gauge unification value.

Throughout this appendix, greek indices refer to gauge eigenstates
while roman indices refer to mass eigenstates for clarity.  This
convention is dropped in later sections to avoid confusion with
Lorentz indices.

\subsection{Higgs Bosons}
Our convention for the Higgs doublets is
\begin{equation}
H_u = \left( \begin{array}{c}
H_u^+ \\
H_u^0 \\
\end{array} \right) 
\qquad 
H_d = \left( \begin{array}{c}
H_d^0 \\
H_d^- \\
\end{array} \right)\, ,
\end{equation}
and we set their vacuum expectation values (VEVs) to be
\begin{equation}
\langle H_u \rangle = \frac{1}{\sqrt{2}} 
\left( \begin{array}{c}
0 \\
v_u \\
\end{array} \right)
\qquad
\langle H_d \rangle = \frac{1}{\sqrt{2}} 
\left( \begin{array}{c}
v_d \\
0 \\
\end{array} \right).
\end{equation}
We set the SM Higgs VEV $v$ to be $v^2 \equiv v_u^2 + v_d^2$, and also
we define $\tan \beta = v_u / v_d$.  Correspondingly, the SM gauge
boson masses are
\begin{equation}
m_W^2 = \frac{1}{4} g^2 v^2 \qquad m_Z^2 = \frac{1}{4} \left( g^2 +
g'^2 \right) v^2\, ,
\end{equation}
and the SM lepton masses are
\begin{equation}
m_{\ell_{ij}} = \frac{1}{\sqrt{2}}
y_{ij}^{(\ell)} v_d\, .
\end{equation}

After electroweak symmetry breaking, the neutral Higgs doublet is
given by
\begin{equation}
\left( \begin{array}{c} 
H_u^0 \\ H_d^0 \\ 
\end{array} \right)
= \frac{1}{\sqrt{2}} 
\left( \begin{array}{c} 
v_u \\ v_d \\
\end{array} \right)
+ \frac{1}{\sqrt{2}} 
R_{\theta_H} 
\left( \begin{array}{c} 
h^0 \\ H^0 \\
\end{array} \right)
+ \frac{i}{\sqrt{2}} 
R_\beta 
\left( \begin{array}{c} 
G^0 \\ A^0 \\
\end{array} \right)\, ,
\end{equation}
and the charged Higgs doublet is 
\begin{equation}
\left( \begin{array}{c} 
H_u^+ \\ H_d^{-*} \\ 
\end{array} \right)
= R_\beta 
\left( \begin{array}{c} 
G^+ \\ H^+ \\ 
\end{array} \right)\, ,
\end{equation}
where 
\begin{equation}
R_{\theta_H} =
\left( \begin{array}{cc}
  \cos \theta_H & \sin \theta_H \\
- \sin \theta_H & \cos \theta_H \\
\end{array} \right)
\quad \text{and} \quad
R_\beta =
\left( \begin{array}{cc}
  \sin \beta & \cos \beta \\
- \cos \beta & \sin \beta \\
\end{array} \right) \ .
\end{equation}

\subsection{Neutralinos}

The neutralino gauge eigenstates are defined by $\tilde{\psi}^0 =
\left( - i \tilde{B}, - i \tilde{W}, \tilde{\psi}_d^0,
\tilde{\psi}_u^0 \right)^T,$ with the ordering of the up/down higgsinos
chosen to conform with {\tt SOFTSUSY}.  All the above fields are
two-component Weyl spinors.

The neutralino mass term in the Lagrangian is given by
\begin{equation}
\mathcal{L} = - \frac{1}{2} \tilde{\psi}_\alpha^{0 \dagger}
M_{\tilde{\psi}^0 \alpha \beta} \tilde{\psi}_\beta^0 \ ,
\end{equation}
so diagonalizing according to $O_{i \alpha}^{\dagger}
M_{\tilde{\psi}^0 \alpha \beta} O_{\beta j} = m_{\chi_i^0}
\delta_{ij}$ gives neutralino mass eigenstates of $\chi_i^0 = O_{i
\alpha}^{\dagger} \tilde{\psi}_\alpha^0$.  The corresponding Dirac
spinors are $\tilde{\chi}_i^0 = \left(\chi_i^0, \overline{\chi}_i^0
\right)^T$.  Here we follow {\tt SOFTSUSY} convention for the
neutralino mass matrix by keeping the matrix real and allowing the
neutralino masses to be negative.

\subsection{Charginos}

Similarly, the chargino gauge eigenstates are defined by
$\tilde{\psi}^+ = \left( - i \tilde{W}^+, \tilde{\psi}_u^+ \right)^T$
and $\tilde{\psi}^- = \left( - i \tilde{W}^-, \tilde{\psi}_d^-
\right)^T$.  Again, these are two-component Weyl spinors.

The Lagrangian mass term is given by
\begin{equation}
\mathcal{L} = - \tilde{\psi}^{- \dagger} M_{\tilde{\psi}^+}
\tilde{\psi}^+\, ,
\end{equation}
so diagonalizing with the rotation angles $\theta_L$ and $\theta_R$
gives
\begin{equation}
\left( \begin{array}{cc}
  \cos \theta_L & \sin \theta_L \\
- \sin \theta_L & \cos \theta_L \\
\end{array} \right)
M_{\tilde{\psi}^+}
\left( \begin{array}{cc}
  \cos \theta_R & - \sin \theta_R \\
  \sin \theta_R &   \cos \theta_R \\
\end{array} \right) =
\left( \begin{array}{cc}
m_{\chi_1^+} & 0 \\
0 & m_{\chi_2^+} \\
\end{array} \right).
\end{equation}
The chargino mass eigenstates thus become
\begin{equation}
\left( \begin{array}{c} \chi_1^+ \\ \chi_2^+ \end{array} \right) =
\left( \begin{array}{cc}
\cos \theta_R & \sin \theta_R \\ - \sin \theta_R & \cos \theta_R
\end{array} \right)
\left( \begin{array}{c} - i \tilde{W}^+ \\ \tilde{\psi}_u^+ 
\end{array} \right)
\end{equation}
and 
\begin{equation}
\left( \begin{array}{c} \chi_1^- \\ \chi_2^- \end{array} \right) =
\left( \begin{array}{cc}
\cos \theta_L & \sin \theta_L \\ - \sin \theta_L & \cos \theta_L
\end{array} \right)
\left( \begin{array}{c} - i \tilde{W}^- \\ \tilde{\psi}_d^- \end{array} 
\right).
\end{equation}
The corresponding Dirac spinors are
$\tilde{\chi}_i = \left( \chi_i^+, \overline{\chi}_i^- \right)^T$.

\subsection{Charged Sleptons}

The mass term for charged sleptons is
\begin{equation}
\mathcal{L} = \tilde{\ell}_\alpha^*
\left(M_{\tilde{\ell}}^2 \right)_{\alpha \beta} \tilde{\ell}_\beta \, .
\end{equation}
Here $\tilde{\ell}_\alpha= \left( \tilde{e}_L, \tilde{\mu}_L,
\tilde{\tau}_L, \tilde{e}_R, \tilde{\mu}_R, \tilde{\tau}_R \right)^T$
is a column vector containing the left-handed sleptons in the first
three components and the right-handed sleptons in the last three
components.

The $6 \times 6$ mass-squared matrix for sleptons is
\begin{equation}
M_{(\tilde{\ell})}^2 = \left(
\begin{array}{cc}
M_L^2 & A \\
A^{\dagger} & M_R^2
\end{array} \right).
\end{equation}
The mass matrix is diagonalized according to $\left(M_{\tilde{\ell}}^2
\right)_{\alpha \beta} = U_{\alpha i}^{(\tilde{\ell})}
\left(M_{\tilde{\ell}}^2 \right)_{ii} U_{i
\beta}^{(\tilde{\ell})\dagger}$, so the slepton mass eigenstates are
defined by $\tilde{\ell}_i = U_{i \alpha}^{(\tilde{\ell})\dagger}
\tilde{\ell}_\alpha$ and $\tilde{\ell}_\alpha = U_{\alpha
i}^{(\tilde{\ell})} \tilde{\ell}_i$, with the mass eigenstates ordered
in terms of increasing mass.

Note that the first index of the matrix $U$ is a gauge index and the
second is a mass index; almost all interactions treat the left- and
right- handed sleptons differently, so typically the first index will
be summed from either 1 to 3 or 4 to 6, while the second will be
summed over all six lepton mass eigenstates.  Note also that in our
convention the rotation matrix is the hermitian conjugate of the
rotation matrix presented in Ref.~\cite{Allanach:2008qq}.

\subsection{Sneutrinos}

The sneutrino mass term is
\begin{equation}
\mathcal{L} = \tilde{\nu}_\alpha^*
\left(M_{\tilde{\nu}}^2 \right)_{\alpha \beta} \tilde{\nu}_\beta\, ,
\end{equation}
with $\tilde{\nu}_\alpha = \left(\tilde{\nu}_e, \tilde{\nu}_\mu,
\tilde{\nu}_\tau \right)^T$.  Diagonalizing gives
$\left(M_{\tilde{\nu}}^2 \right)_{\alpha \beta} = U_{\alpha
i}^{(\tilde{\nu})} \left(M_{\tilde{\nu}}^2 \right)_{ii} U_{i
\beta}^{(\tilde{\nu})\dagger}$, and thus $\tilde{\nu}_i = U_{i
\alpha}^{(\tilde{\nu})\dagger} \tilde{\nu}_\alpha$ and
$\tilde{\nu}_\alpha = U_{\alpha i}^{(\tilde{\nu})} \tilde{\nu}_i$.
Again our rotation matrix is the hermitian conjugate of the one in
\cite{Allanach:2008qq}.

\subsection{Leptons}

As previously mentioned, we work in the basis of charged leptons in
which the mass and flavor eigenstates are in correspondence.  This is
possible because there is a freedom of a single superfield rotation,
and we choose to use this rotation to rotate away any mixing in the
charged lepton sector (this was done for both notational simplicity
and convenient coding).  Also, as mentioned previously, we have
neglected neutrino mixing, as neutrino flavor is not observable at
colliders.


\section{Lagrangians}
\label{app:lagrangians}

Indices in Lagrangian coefficients are ordered by ``gaugino, sfermion,
fermion'' if there is a gaugino present.  Our Lagrangian assumes no
neutrino masses and thus that the charged leptons can freely be
rotated into mass eigenstates.  Thus all the mixing terms will be
parameterized by only the slepton and sneutrino mixing matrices (and
the standard neutralino/chargino mixing matrices).

For the mixing matrices, the left and right bases are different, with
one being the gauge basis and the other the mass basis.  We have
chosen the first index to correspond to a gauge eigenstate and the
second to correspond to a mass eigenstate.  Thus, in a term like
$U_{c, b}^{(\tilde{\ell})}$, $c$ is the gauge index and $b$ is a
slepton mass index.  Note that in summations over these indices, the
gauge summations are always taken from 1 to 3 and the mass summations
are taken from 1 to 6 for sleptons and 1 to 3 for sneutrinos.

In this section, we only consider phenomenologically viable decays to
lowest order; thus we neglect four scalar couplings and couplings for
modes that are unlikely to be kinematically allowed.

\subsection{Neutralino Terms}
The neutralino interaction terms are
\begin{equation}
\mathcal{L}_{\text{Neutralino}} = \left[
\alpha_{abc} \tilde{\nu}_b^* \overline{\tilde{\chi}}_a^0 P_L \nu_c
+ \tilde{\ell}_b^* \overline{\tilde{\chi}}_a^0
\left( \beta_{abc}^{(1)}  P_L + \beta_{abc}^{(2)} P_R \right)
\ell_c \right] + \text{ h.c.}\, ,
\end{equation}
where
\begin{eqnarray}
\alpha_{abc} &=& - \frac{1}{\sqrt{2}}
\left( g O_{2, a}^* - g' O_{1, a}^* \right) U_{c, b}^{(\tilde{\nu})*} \\
\beta_{abc}^{(1)} &=& \frac{1}{\sqrt{2}}
\left( g O_{2, a}^* + g' O_{1, a}^* \right) U_{c, b}^{(\tilde{\ell})*} 
- y_{c}^{(\ell)} O_{3, a}^* U_{c + 3, b}^{(\tilde{\ell})*} \\
\beta_{abc}^{(2)} &=& - \sqrt{2} O_{1, a} g' U_{c + 3, b}^{(\tilde{\ell})*}
- y_{c}^{(\ell)} O_{3, a} U_{c, b}^{(\tilde{\ell})*} .
\end{eqnarray}
For these terms, the matrix $O_{i, j}$ is the neutralino mixing matrix
and $y_{c}^{(\ell)}$ is the $c$ component of the diagonal lepton
Yukawa matrix; the gauge components correspond to 1 - bino, 2 - wino,
3 - down Higgsino\footnote{The up Higgsino corresponds to $a = 4$, but
this is never needed.}.

\subsection{Chargino Terms}
The chargino interaction terms are
\begin{equation}
\mathcal{L}_{\text{Chargino}} = \left\{
\gamma_{abc} \tilde{\ell}_b^* \overline{\tilde{\chi}}_a P_L \nu_c
+ \tilde{\nu}^*_b \overline{\tilde{\chi}_a^c} \left( \delta_{abc}^{(1)} P_L
+ \delta_{abc}^{(2)} P_R \right) \ell_c \right\} + \text{ h.c.},
\end{equation}
where
\begin{eqnarray}
\gamma_{abc} &=& \left\{ \begin{array}{ll}
-g \cos \theta_L \, U_{c, b}^{(\tilde{\ell})*} +
y_{c}^{(\ell)} \sin \theta_L \, U_{c + 3, b}^{(\tilde{\ell})*}
& \qquad a = 1 \\
g \sin \theta_L \, U_{c, b}^{(\tilde{\ell})*} +
y_{c}^{(\ell)} \cos \theta_L \, U_{c + 3, b}^{(\tilde{\ell})*}\,
& \qquad a = 2 \\
\end{array} \right. \\
\delta_{abc}^{(1)} &=& \left\{ \begin{array}{ll}
- g \cos \theta_R \, U_{c, b}^{(\tilde{\nu})*} & \qquad a = 1 \\
g \sin \theta_R \, U_{c, b}^{(\tilde{\nu})*}\, & \qquad a = 2 \\
\end{array} \right. \\
\delta_{abc}^{(2)} &=& \left\{ \begin{array}{ll}
y_{c}^{(\ell)} \sin \theta_L \, U_{c, b}^{(\tilde{\nu})*} & \qquad a = 1 \\
y_{c}^{(\ell)} \cos \theta_L \, U_{c, b}^{(\tilde{\nu})*} & \qquad a = 2 \\
\end{array} \right. \, .
\end{eqnarray}
Here $\theta_L$ and $\theta_R$ are the rotation angles associated with
rotating the negatively and positively charged (respectively) gauginos
and higginos into the mass eigenstates.

\subsection{Higgs Terms}
The neutral Higgs interaction terms are
\begin{equation}
\mathcal{L}_{\text{Neutral Higgs}} = 
  \sigma_{ab}^{(1)} \tilde{\nu}_a^* \tilde{\nu}_b H^0
+ \sigma_{ab}^{(2)} \tilde{\ell}_a^* \tilde{\ell}_b h^0
+ \sigma_{ab}^{(3)} \tilde{\ell}_a^* \tilde{\ell}_b H^0
+ i \sigma_{ab}^{(4)} \tilde{\ell}_a^* \tilde{\ell}_b A^0
+ i \sigma_{ab}^{(5)} \tilde{\ell}_a^* \tilde{\ell}_b G^0.
\end{equation}
There are five different $\sigma$ coefficients since the two slepton
couplings to $h^0$ and $H^0$ are different:

\begin{eqnarray}
\sigma_{ab}^{(1)} &=& - \frac{g m_W}{
2 \cos^2 \theta_W} \cos (\theta_H + \beta) \delta_{ab} \\
\nonumber \sigma_{ab}^{(2)} &=& - \left[ \left( \frac{g m_W}{2} (1 -
  \tan^2 \theta_W) \sin (\theta_H + \beta) - \frac{g m_{\ell_c}^2 \sin
    \theta_H}{ m_W \cos \beta} \right) U_{c,a}^{(\tilde{\ell})*}
  U_{c,b}^{(\tilde{\ell})} \right. \\
& & \quad + \left( g m_W \tan^2 \theta_W \sin (\theta_H + \beta) -
\frac{g m_{\ell_c}^2 \sin \theta_H}{ m_W \cos \beta} \right)
U_{c+3,a}^{(\tilde{\ell})*} U_{c+3,b}^{(\tilde{\ell})} \nonumber \\
& & \quad \left.  - \frac{g m_{\ell_c}}{2 m_W \cos \beta} \left(\mu
\cos \theta_H + A_c^{\tilde{\ell}} \sin \theta_H \right)
\left (U_{c,a}^{(\tilde{\ell})*} U_{c+3,b}^{(\tilde{\ell})} +
U_{c+3,a}^{(\tilde{\ell})*} U_{c,b}^{(\tilde{\ell})}\right) \right] \\
\nonumber \sigma_{ab}^{(3)} &=& \left[ \left( \frac{g m_W}{2} (1 -
\tan^2 \theta_W) \cos (\theta_H + \beta) - \frac{g m_{\ell_c}^2 \cos
\theta_H}{ m_W \cos \beta} \right) U_{c,a}^{(\tilde{\ell})*}
U_{c,b}^{(\tilde{\ell})} \right. \\
& & \quad + \left( g m_W \tan^2 \theta_W \cos (\theta_H + \beta) -
\frac{g m_{\ell_c}^2 \cos \theta_H}{ m_W \cos \beta} \right)
U_{c+3,a}^{(\tilde{\ell})*} U_{c+3,b}^{(\tilde{\ell})} \nonumber \\
& &\quad \left.  + \frac{g m_{\ell_c}}{2 m_W \cos \beta} \left(\mu
\sin \theta_H - A_c^{\tilde{\ell}} \cos \theta_H \right)
\left(U_{c,a}^{(\tilde{\ell})*} U_{c+3,b}^{(\tilde{\ell})} +
U_{c+3,a}^{(\tilde{\ell})*} U_{c,b}^{(\tilde{\ell})}\right) \right] \\
\sigma_{ab}^{(4)} &=& \frac{g m_{\ell_c}}{ 2 m_W} (\mu +
A_c^{\tilde{\ell}} \tan \beta) \left( U_{c, a}^{(\tilde{\ell})*} U_{c
+ 3, b}^{(\tilde{\ell})} - U_{c + 3, a}^{(\tilde{\ell})*} U_{c,
b}^{(\tilde{\ell})} \right) \\
\sigma_{ab}^{(5)} &=& \frac{g m_{\ell_c}}{ 2 m_W} (\mu \tan \beta +
A_c^{\tilde{\ell}}) \left( U_{c, a}^{(\tilde{\ell})*} U_{c + 3,
b}^{(\tilde{\ell})} - U_{c + 3, a}^{(\tilde{\ell})*} U_{c,
b}^{(\tilde{\ell})} \right).
\end{eqnarray}
Here $\tan \theta_W = g' / g$ as usual, $\tan \beta = v_u / v_d$, and
the angle $\theta_H$ is the rotation angle for the neutral Higgs
scalars $H_u$ and $H_d$ into the real neutral Higgs mass eigenstates.
The light and heavy real Higgs bosons are denoted $h^0$ and $H^0$, and
the pseudoscalar Higgs is denoted $A^0$.  The parameter $\mu$ is the
Higgsino mass parameter, and $A_{\ell_i}$ is the left-right mixing
term for $\ell_i$ in the diagonal lepton basis.

It is useful to note that several of the neutral Higgs terms are proportional
to $U_{c, a}^{(\tilde{\ell})*} U_{c, b}^{(\tilde{\ell})}$,
$U_{c+3, a}^{(\tilde{\ell})*} U_{c+3, b}^{(\tilde{\ell})}$, or
$U_{c+3, a}^{(\tilde{\ell})*} U_{c, b}^{(\tilde{\ell})}$.  The first two are
``block flavor diagonal'' in left and right sleptons and the third has explicit
left-right mixing, so flavor mixing decays of these modes should be
significantly suppressed relative to chargino/neutralino modes by the size of
left-right mixing, as discussed in Ref.~\cite{Feng:3body}.

Using the conventions above, the charged Higgs interaction terms are
\begin{equation}
\mathcal{L}_{\text{Charged Higgs}} =
\left[ \rho_{ab}^{(1)} \tilde{\nu}_a^* \tilde{\ell}_b H^+ +
\rho_{ab}^{(2)} \tilde{\nu}_a^* \tilde{\ell}_b G^+ \right] + \text{ h.c.}\, ,
\end{equation}
where
\begin{eqnarray}
\nonumber \rho_{ab}^{(1)} & = & - g \left[ \left( \frac{m_W}{\sqrt{2}}
  \sin 2 \beta - \frac{m_{\ell_c}^2 \tan \beta}{\sqrt{2} m_W} \right)
  U_{c, a}^{(\tilde{\nu})*} U_{c, b}^{(\tilde{\ell})} \right. \\
& & \left. - \frac{m_{\ell_c}}{\sqrt{2} m_W} (\mu + A_{\ell_c}\tan
  \beta) U_{c, a}^{(\tilde{\nu})*} U_{c + 3, b}^{(\tilde{\ell})}
  \right] \\
\nonumber \rho_{ab}^{(2)} & = & g \left[ \left( \frac{m_W}{\sqrt{2}}
\cos 2 \beta - \frac{m_{\ell_c}^2}{\sqrt{2} m_W} \right) U_{c,
a}^{(\tilde{\nu})*} U_{c, b}^{(\tilde{\ell})} \right. \\
& & \left. + \frac{m_{\ell_c}}{\sqrt{2} m_W} (\mu \tan \beta -
  A_{\ell_c}) U_{c, a}^{(\tilde{\nu})*} U_{c + 3, b}^{(\tilde{\ell})}
  \right].
\end{eqnarray}
Note there is no additional rotation angle since $\beta$ serves as the
rotation angle for the charged Higgs.

\subsection{Gauge Bosons}

The Lagrangian interaction terms for the gauge bosons are
\begin{eqnarray}
\nonumber L_{\text{Gauge Boson}} & = & \left[ i \zeta^{(1)}_{ab} \left(
\tilde{\nu}_a^* \partial_{\mu} \tilde{\ell}_b - \tilde{\ell}_b
\partial_{\mu} \tilde{\nu}_a^* \right) W^{+ \mu} + \text{ h.c.}
\right] \\
& & + i \zeta^{(2)}_{ab} \left(\tilde{\ell}_a^* \partial_{\mu}
\tilde{\ell}_b - \tilde{\ell}_b \partial_{\mu} \tilde{\ell}_a^*
\right) Z^{\mu} + i \zeta^{(3)}_{ab} \left(\tilde{\ell}_a^*
\partial_{\mu} \tilde{\ell}_b - \tilde{\ell}_b \partial_{\mu}
\tilde{\ell}_a^* \right) A^{\mu} \, ,
\end{eqnarray}
where
\begin{eqnarray}
\zeta^{(1)}_{ab} & = & -\frac{g}{\sqrt{2}} 
U_{c, a}^{(\tilde{\nu})*} U_{c, b}^{(\tilde{\ell})} \\
\zeta^{(2)}_{ab} & = & \frac{g}{2 \cos \theta_W} 
\left[ U_{c, a}^{(\tilde{\ell})*} U_{c, b}^{(\tilde{\ell})}
- 2 \sin^2 \theta_W \delta_{ab} \right] \\
\zeta^{(3)}_{ab} & = & e \delta_{ab} \, .
\end{eqnarray}
Here, $e$ is the electromagnetic charge.  As above for neutral Higgs
bosons, the flavor mixing modes involving $Z$ bosons are heavily
suppressed by the left-right mixing.  The photon vertex remains flavor
diagonal even in the flavor mixing case, as expected.


\section{Table of Decay Modes}
\label{app:decaytable}

\begin{center}
\begin{tabular}{|c|l|l|}
\hline
{\bf Decay Number} & {\bf Decay Mode} & {\bf Corresponding Program Method} \\
\hline
1 & $\tilde{\chi}_i^0 \to \tilde{\ell}_j^- \ell_k^+$ &
{\tt CalcChi0SlepLepBar} \\
2 & $\tilde{\chi}_i^0 \to \tilde{\nu}_j \overline{\nu}_k$ &
{\tt CalcChi0SnuNuBar} \\
3 & $\tilde{\chi}_i^- \to \tilde{\ell}_j^- \overline{\nu}_k$ &
{\tt CalcChiMinusSlepNuBar} \\
4 & $\tilde{\chi}_i^+ \to \tilde{\nu}_j \ell_k^+$ &
{\tt CalcChiPlusSnuLepBar} \\
5 & $\tilde{\ell}_i^- \to \tilde{\chi}_j^0 \ell_k^-$ &
{\tt CalcSlepChi0Lep} \\
6 & $\tilde{\ell}_i^- \to \tilde{\chi}_j^- \nu_k$ &
{\tt CalcSlepChiMinusNu} \\
7 & $\tilde{\ell}_i^- \to \tilde{\ell}_j^- h^0$ &
{\tt CalcSlepSlepHiggs} \\
8 & $\tilde{\ell}_i^- \to \tilde{\ell}_j^- Z^0$ &
{\tt CalcSlepSlepZ} \\
9 & $\tilde{\ell}_i^- \to \tilde{\nu}_j W^-$ &
{\tt CalcSlepSnuWMinus} \\
10 & $\tilde{\nu}_i \to \tilde{\chi}_j^0 \nu_k$ &
{\tt CalcSnuChi0Nu} \\
11 & $\tilde{\nu}_i \to \tilde{\chi}_j^+ \ell_k^-$ &
{\tt CalcSnuChiPlusLep} \\
12 & $\tilde{\nu}_i \to \tilde{\ell}_j^- W^+$ &
{\tt CalcSnuSlepWPlus} \\
13 & $H^0 \to \tilde{\ell}_i^- \tilde{\ell}_j^+$ &
{\tt CalcHSlepSlepBar} \\
14 & $H^0 \to \tilde{\nu}_i \tilde{\nu}_j$ &
{\tt CalcHSnuSnuBar} \\
15 & $A^0 \to \tilde{\ell}_i^- \tilde{\ell}_j^+$ &
{\tt CalcASlepSlepBar} \\
16 & $H^- \to \tilde{\ell}_i^- \tilde{\nu}_j$ &
{\tt CalcHMinusSlepSnuBar} \\
17 & $\tilde{\ell}_i^- \to \tilde{\ell}_j^- \ell_k^- \ell_m^+$ &
{\tt CalcSlepSlepLepLepBar} \\
18 & $\tilde{\ell}_i^- \to \tilde{\ell}_j^+ \ell_k^- \ell_m^-$ &
{\tt CalcSlepSlepBarLepLep} \\
19 & $\tilde{\ell}_i^- \to \tilde{\ell}_j^+ \nu_k \overline{\nu}_m$ &
{\tt CalcSlepSlepNuNuBar} \\
20 & $\tilde{\ell}_i^- \to \tilde{\ell}_j^+ q_k \overline{q}_k$ &
{\tt CalcSlepSlepQQBar} \\
\hline
\end{tabular} 
\vspace*{0.1in} \newline
Table II: List of included decay modes with corresponding program
methods in order of appearance.
\end{center}

\begin{center}
\begin{tabular}{|l|l|}
\hline
{\bf Neglected Decay Modes} & {\bf Justification} \\
\hline
$\tilde{\ell}_i^- \to \tilde{\ell}_j^- H^0$ &
Kinematics --- $H^0$ assumed heavier than $\tilde{\ell}$ \\
$\tilde{\ell}_i^- \to \tilde{\ell}_j^- A^0$ &
Kinematics --- $A^0$ assumed heavier than $\tilde{\ell}$ \\
$\tilde{\ell}_i^- \to \tilde{\nu}_j H^-$ &
Kinematics --- $H^-$ assumed heavier than $\tilde{\ell}$ \\
$\tilde{\nu}_i \to \tilde{\nu}_j Z$ &
Kinematics --- $Z$ assumed heavier than $\tilde{\nu}$ splitting \\
$\tilde{\nu}_i \to \tilde{\nu}_j h^0$ &
Kinematics --- $h^0$ assumed heavier than $\tilde{\nu}$ splitting \\
$\tilde{\nu}_i \to \tilde{\nu}_j H^0$ &
Kinematics --- $H^0$ assumed heavier than $\tilde{\nu}$ \\
$\tilde{\nu}_i \to \tilde{\ell}_j^- H^+$ &
Kinematics --- $H^+$ assumed heavier than $\tilde{\nu}$ \\
$h^0 \to \tilde{\ell}_i^- \tilde{\ell}_j^+$ &
Kinematics --- multiple $\tilde{\ell}$'s assumed heavier than $h^0$ \\
$h^0 \to \tilde{\nu}_i^- \tilde{\nu}_j^+$ &
Kinematics --- multiple $\tilde{\nu}$'s assumed heavier than $h^0$ \\
Any with emitted $\gamma$ & QED interactions still flavor neutral \\
Scalar three-body decays & Suppression by couplings and phase space factors \\
\hline
\end{tabular} 
\vspace*{0.1in} \newline
Table III: List of neglected decay modes, and reason for omission.
\end{center}


\section{Two Body Decays}
\label{app:twobody}

This section contains the matrix elements and decay widths for all
two-body decays.  Charge conjugate modes are not included; we work in
a non-CP violating lepton sector, so charge conjugate modes in all
cases have the same decay widths as the original modes.  Again,
neutrinos are taken to be massless.

Our index notation is that the index $i$ refers to the decaying
particle and the indices $j$ and $k$ to the decay products.  Further,
if the decaying particle is a slepton, sneutrino, chargino, or
neutralino then the index $j$ refers to the resultant SUSY particle
while $k$ refers to the resultant Standard Model particle; this
convention does not apply to Higgs decays, which have two SUSY
daughters.

The general formula for the decay width of a two-body decay in the
center of mass frame of the decaying $i$ particle with daughters $j$
and $k$ is
\begin{equation}
\Gamma_{\text{2 body decay}} = \frac{| \mathcal{M} |^2}{16 \pi m_i^3}
\lambda^{1/2} \left( m_i^2, m_j^2, m_k^2 \right)\, ,
\end{equation}
where $\lambda(x, y, z) = x^2 + y^2 + z^2 - 2(xy + xz + yz)$, and the
matrix element squared has an implicit average/sum over fermion spins
if the decay involves fermions.  For our purposes, there are three
general cases of matrix element forms: those containing two fermions
(lepton and neutralino/chargino), those containing gauge bosons, and
those containing only sleptons and Higgs scalars.  The scalar case is
trivial, since the matrix element is equal to the Lagrangian term with
no momentum dependence, and the decay width is simply the square of
the Lagragian term multiplied by a phase space factor.  For fermionic
modes with fermions $f_1$ and $f_2$ and a scalar $s$, the matrix
element takes the form of
\begin{equation}
\mathcal{M}_{\text{fermionic 2-body}} \sim i \left( \begin{array}{c}
\overline{u}_1 (p_{f_1}) \\ 
\overline{v}_1 (p_{f_1}) \\ 
\end{array} \right)
\left( A P_R + B P_L \right) \left( \begin{array}{c} 
u_2 (p_{f_2}) \\ v_2 (p_{f_2}) \\ 
\end{array} \right) \, ,
\end{equation}
which gives a decay width proportional to $\pm ( | A |^2 + | B |^2 ) (
m_{f_1}^2 + m_{f_2}^2 - m_{s}^2 ) \pm 4 m_{f_1} m_{f_2} \text{ Re}
\left( A B^* \right)$ times the standard phase space factor (with the
possibility of $m_{f_2} = 0$ if a neutrino is a decay product).  The
sign on the second term will be positive if both spinors are $u$ or
$v$, and negative if they are different spinors.

Finally, gauge boson matrix elements with scalars $s_1$ and $s_2$ and
a gauge boson $b$ are proportional to $(p_{s_1} + p_{s_2})^\mu
\epsilon_\mu$, so the decay width will be proportional to
$\lambda^{\frac{1}{2}} \left(m_{s_1}^2,m_{s_2}^2,m_{b}^2
\right)/m_{b}^2$.

\subsection{Neutralino Decays}
For a neutralino two-body decay to charged slepton and lepton, the
flavor-generalized matrix element is
\begin{equation}
\mathcal{M} \left( \tilde{\chi}_i^0 \to \tilde{\ell}_j^- \ell_k^+
\right) = i \overline{u}_{\ell_k} (p_k) \left(\beta_{ijk}^{(1)*} P_R +
\beta_{ijk}^{(2)*} P_L \right) u_{\tilde{\chi}_i^0}(p_i)\, ,
\end{equation}
and the decay width is
\begin{eqnarray}
\nonumber \lefteqn{\Gamma \left( \tilde{\chi}_i^0 \to \tilde{\ell}_j^-
\ell_k^+ \right) = \frac{1}{32 \pi m_{\tilde{\chi}_i^0}^3}
\lambda^{1/2} \left(m_{\tilde{\chi}_i^0}^2, m_{\tilde{\ell}_j}^2,
m_{\ell_k}^2\right)} \\
& \left[ \left( \left| \beta_{ijk}^{(1)} \right|^2 + \left|
\beta_{ijk}^{(2)} \right|^2 \right) \left( m_{\tilde{\chi}_i^0}^2 +
m_{\ell_k^+}^2 - m_{\tilde{\ell}_j^-}^2 \right) + 4
m_{\tilde{\chi}_i^0} m_{\ell_k^+} \text{ Re} \left( \beta_{ijk}^{(1)*}
\beta_{ijk}^{(2)} \right) \right].
\end{eqnarray}
For a neutralino two-body decay to sneutrino and neutrino, the matrix
element is
\begin{equation}
\mathcal{M} (\tilde{\chi}_i^0 \to \tilde{\nu}_{\ell_j}
\overline{\nu}_k) = i \alpha_{ijk}^* \overline{v}_{\nu_k} (p_k) P_R
v_{\tilde{\chi}_i^0}(p_i)\, ,
\end{equation}
and the decay width is
\begin{equation}
\Gamma (\tilde{\chi}_i^0 \to \tilde{\nu}_{\ell_j} \overline{\nu}_k) =
\frac{1}{32 \pi m_{\tilde{\chi}_i^0}^3} \left| \alpha_{ijk} \right|^2
\left( m_{\tilde{\chi}_i^0}^2 - m_{\tilde{\nu}_j}^2 \right)^2 \ .
\end{equation}

\subsection{Chargino Decays}
The matrix element for a chargino decay to slepton and neutrino is
\begin{equation}
\mathcal{M} \left( \tilde{\chi}_i^- \to \tilde{\ell}_j^-
\overline{\nu}_{\ell_k} \right) = \gamma_{ijk}^*
\overline{v}_{\nu_{\ell_k}} (p_k) P_R v_{\tilde{\chi}_i} (p_i) \, ,
\end{equation}
and the decay width is
\begin{equation}
\Gamma \left( \tilde{\chi}_i^- \to \tilde{\ell}_j^-
\overline{\nu}_{\ell_k} \right) = \frac{|\gamma_{ijk}|^2}
{32 \pi m_{\tilde{\chi}_i}^3}
\left( m_{\tilde{\chi}_i}^2 - m_{\tilde{\ell}_j}^2 \right)^2.
\end{equation}

The matrix element for a chargino decay to sneutrino and lepton is
\begin{eqnarray}
\nonumber \mathcal{M} \left(\tilde{\chi}_i \to \tilde{\nu}_j \ell_k^+
\right) & = & i \overline{v}_{\ell_k} (p_k) \left(\delta_{jik}^{(1)*}
P_R + \delta_{jik}^{(2)*} P_L \right) \mathcal{C}
\left[\overline{u}_{\tilde{\chi}_i} (p_i) \right]^T \\
& = & i \overline{v}_{\ell_k} (p_k) \left(\delta_{jik}^{(1)*} P_R +
  \delta_{jik}^{(2)*} P_L \right) v_{\tilde{\chi}_i} (p_j)\, ,
\end{eqnarray}
and the decay width is
\begin{eqnarray}
\nonumber \lefteqn{\Gamma \left(\tilde{\chi}_i \to \tilde{\nu}_j
\ell_k^+ \right) = \frac{1}{32 \pi m_{\tilde{\chi}_i}^3} \lambda^{1/2}
\left(m_{\tilde{\chi}_i}^2, m_{\tilde{\nu}_j}^2, m_{\ell_k}^2\right)}
\\
& \left[ \left( \left| \delta_{ijk}^{(1)} \right|^2 + \left|
  \delta_{ijk}^{(2)} \right|^2 \right) \left( m_{\tilde{\chi}_i}^2 +
  m_{\ell_k^-}^2 - m_{\tilde{\nu}_j}^2 \right) + 4 m_{\tilde{\chi}_i}
  m_{\ell_k} \text{ Re} \left( \delta_{ijk}^{(1)*} \delta_{ijk}^{(2)}
  \right) \right] .
\end{eqnarray}
Note that $\mathcal{C}$ is the charge conjugation operator.

\subsection{Slepton Decays}
The slepton decay to neutralino and lepton has a matrix element given
by
\begin{equation}
\mathcal{M} (\tilde{\ell}_i^- \to \tilde{\chi}_j^0 \ell_k^- ) =
i \overline{u}_{\ell_k} (p_k) \left( \beta_{jik}^{(1)} P_R
  + \beta_{jik}^{(2)} P_L \right) v_{\tilde{\chi}_j^0}(p_j)\, ,
\end{equation}
and the corresponding decay width is
\begin{eqnarray}
\nonumber \lefteqn{\Gamma (\tilde{\ell}_i^- \to \tilde{\chi}_j^0
\ell_k^- ) = \frac{1}{16 \pi m_{\tilde{\ell}_i}^3} \lambda^{1/2}
\left( m_{\tilde{\ell}_i}^2, m_{\tilde{\chi}_j^0}^2, m_{\ell_k}^2
\right)} \\
& \left[ \left( |\beta_{jik}^{(1)}|^2 + |\beta_{jik}^{(2)}|^2 \right)
\left( m_{\tilde{\ell}_i}^2 - m_{\tilde{\chi}_j^0}^2 - m_{\ell_k}^2
\right) - 4 m_{\tilde{\chi}_j^0} m_{\ell_k} \text{ Re} \left(
\beta_{jik}^{(1)*} \beta_{jik}^{(2)} \right) \right].
\end{eqnarray}
For a slepton decay to chargino and neutrino, the matrix element is
\begin{equation}
\mathcal{M} (\tilde{\ell}_{i}^- \to \tilde{\chi}_j^- \nu_{\ell_k}) = i
 \gamma_{jik} \overline{u}_{\nu_k} (p_k) P_R v_{\tilde{\chi}_j^-}
 (p_j)\, ,
\end{equation}
and the decay width is
\begin{equation}
\Gamma (\tilde{\ell}_{i}^- \to \tilde{\chi}_j^- \nu_{\ell_k}) =
\frac{|\gamma_{jik}|^2}{
16 \pi m_{\tilde{\ell}_i}^3}
\left( m_{\tilde{\ell}_i}^2 - m_{\tilde{\chi}_j^-}^2 \right)^2 \, .
\end{equation}
A slepton decay to light (neutral) higgs and slepton has a matrix
element of
\begin{equation}
\mathcal{M} (\tilde{\ell}_i^- \to \tilde{\ell}_j^- h^0) = i
\sigma_{ji}^{(2)}\, ,
\end{equation}
and the decay width is
\begin{equation}
\Gamma (\tilde{\ell}_i^- \to \tilde{\ell}_j^- h^0) =
\frac{|\sigma_{ji}^{(2)}|^2}{ 16 \pi m_{\tilde{\ell}_i}^3}
\lambda^{1/2} \left(m_{\tilde{\ell}_i}^2, m_{\tilde{\ell}_j}^2,
m_h^2\right) \, .
\end{equation}
The slepton to slepton and $Z$ boson matrix element is
\begin{equation}
\mathcal{M} (\tilde{\ell}_i^- \to \tilde{\ell}_j^- Z^0) =
i \zeta_{ji}^{(2)} (p_i + p_j)^{\mu} \epsilon_{\mu}^* (p_Z)\, ,
\end{equation}
and the decay width is
\begin{equation}
\Gamma (\tilde{\ell}_i^- \to \tilde{\ell}_j^- Z^0) = \frac{|
\zeta_{ji}^{(2)} |^2}{ 16 \pi m_Z^2 m_{\tilde{\ell}_i}^3}
\lambda^{3/2} \left(m_{\tilde{\ell}_i}^2, m_{\tilde{\ell}_j}^2,m_Z^2
\right) \, .
\end{equation}
The slepton to sneutrino and $W$ boson matrix element is
\begin{equation}
\mathcal{M} (\tilde{\ell}_i^- \to \tilde{\nu}_j W^-) =
i \zeta_{ji}^{(1)} (p_i + p_j)^{\mu} \epsilon_{\mu}^* (p_W) \, ,
\end{equation}
and the decay width is
\begin{equation}
\Gamma (\tilde{\ell}_i^- \to \tilde{\nu}_j W^-) = \frac{|
\zeta_{ji}^{(1)} |^2}{16 \pi m_W^2 m_{\tilde{\ell}_i}^3} \lambda^{3/2}
\left(m_{\tilde{\ell}_i}^2, m_{\tilde{\nu}_j}^2, m_W^2 \right) \, .
\end{equation}

\subsection{Sneutrino Decays}
The two-body mode for sneutrino decaying to neutralino and neutrino
has a matrix element of
\begin{equation}
\mathcal{M} (\tilde{\nu}_i \to \tilde{\chi}_j^0 \nu_k) = i
 \alpha_{jik} \overline{u}_{\nu_k} (p_k) P_R v_{\tilde{\chi}_j^0}
 (p_j)\, ,
\end{equation}
and a decay width 
\begin{equation}
\Gamma(\tilde{\nu}_i \to \tilde{\chi}_j^0 \nu_k) =
 \frac{|\alpha_{jik}|^2 }{
16 \pi m_{\tilde{\nu}_i}^3}
\left( m_{\tilde{\nu}_i}^2 - m_{\tilde{\chi}_j^0}^2 \right)^2 \, .
\end{equation}

The sneutrino to chargino and lepton two-body mode has a matrix element of
\begin{eqnarray}
\nonumber \mathcal{M} \left(\tilde{\nu}_i \to \tilde{\chi}_j^+
\ell_k^- \right) & = & i \overline{u}_{\ell_k^-} (p_k)
\left(\delta_{jik}^{(1)} P_R + \delta_{jik}^{(2)} P_L \right)
\mathcal{C} \left[\overline{v}_{\tilde{\chi}_j^+} (p_i) \right]^T \\
& = & i \overline{u}_{\ell_k^-} (p_k) \left(\delta_{jik}^{(1)} P_R +
  \delta_{jik}^{(2)} P_L \right) u_{\tilde{\chi}_j^+} (p_j)\, ,
\end{eqnarray}
and a decay width of
\begin{eqnarray}
\nonumber \lefteqn{\Gamma \left(\tilde{\nu}_i \to \tilde{\chi}_j^+
\ell_k^- \right) = \frac{1}{16 \pi m_{\tilde{\nu}_i}^3} \lambda^{1/2}
\left(m_{\tilde{\nu}_i}^2, m_{\tilde{\chi}_j^+}^2, m_{\ell_k^-}^2
\right)} \\
& \left[ \left( \left| \delta_{jik}^{(1)} \right|^2 + \left|
\delta_{jik}^{(2)} \right|^2 \right) \left( m_{\tilde{\nu}_i}^2 -
m_{\tilde{\chi}_j^+}^2 - m_{\ell_k}^2 \right) + 4 m_{\tilde{\chi}_j^+}
m_{\ell_k} \text{ Re} \left( \delta_{jik}^{(1)*} \delta_{jik}^{(2)}
\right) \right].
\end{eqnarray}

The matrix element for sneutrinos decaying to sleptons and $W$ bosons is
\begin{equation}
\mathcal{M} (\tilde{\nu}_i \to \tilde{\ell}_j^- W^+) =
i \zeta_{ij}^{(1)} (p_i + p_j)^{\mu} \epsilon_{\mu}^* (p_W)\, ,
\end{equation}
and the decay width is
\begin{equation}
\Gamma (\tilde{\nu}_i \to \tilde{\ell}_j^- W^+) = \frac{|
\zeta_{ij}^{(1)} |^2}{16 \pi m_W^2 m_{\tilde{\nu}_i}^3} \lambda^{3/2}
\left(m_{\tilde{\nu}_i}^2, m_{\tilde{\ell}_j}^2, m_W^2 \right).
\end{equation}

\subsection{Higgs Decays}
Finally, the Higgs decays have trivial matrix elements, so we only
present their widths.  The decay width for heavy Higgs boson to two
sleptons is
\begin{equation}
\Gamma (H^0 \to \tilde{\ell}_j^- \tilde{\ell}_k^+) =
 \frac{|\sigma_{jk}^{(3)}|^2}{ 16 \pi m_H^3} \lambda^{1/2}
 \left(m_H^2, m_{\tilde{\ell}_j}^2, m_{\tilde{\ell}_k}^2 \right) \, ,
\end{equation}
while the width to two sneutrinos is
\begin{equation}
\Gamma (H^0 \to \tilde{\nu}_j \tilde{\nu}_k) =
 \frac{|\sigma_{jk}^{(1)}|^2}{ 16 \pi m_H^3} \lambda^{1/2}
 \left(m_H^2, m_{\tilde{\nu}_j}^2, m_{\tilde{\nu}_k}^2 \right).
\end{equation}

The decay width for pseudoscalar Higgs to two sleptons is
\begin{equation}
\Gamma (A^0 \to \tilde{\ell}_j^- \tilde{\ell}_k^+) =
\frac{|\sigma_{jk}^{(4)}|^2}{ 16 \pi m_{A^0}^3} \lambda^{1/2}
\left(m_{A^0}^2, m_{\tilde{\ell}_j}^2, m_{\tilde{\ell}_k}^2 \right)\,
,
\end{equation}
and the decay width for charged Higgs to slepton and sneutrino is
\begin{equation}
\Gamma (H^- \to \tilde{\ell}_j^- \tilde{\nu}_k) =
\frac{|\rho_{jk}|^2}{ 16 \pi m_{H^+}^3} \lambda^{1/2} \left(m_{H^+}^2,
m_{\tilde{\nu}_j}^2, m_{\tilde{\ell}_k}^2 \right) \ .
\end{equation}


\section{Three Body Decays}
\label{app:threebody}

The general three-body decay width is
\begin{equation}
\Gamma_{\text{3 body decay}} = \frac{1}{64 \pi^3 m_i}
\int_{m_k}^{\frac{1}{2m_i}(m_i^2 + m_k^2 - (m_j + m_m)^2)} dE_k
\int_{E_m^-}^{E_m^+} dE_m \left| \mathcal{M} \right|^2 \, ,
\end{equation}
where the squared matrix element includes a sum over the outgoing
fermion spins.  The limits are
\begin{eqnarray}
\nonumber & E_m^\pm = \frac{1}{2 \left(p_i - p_k \right)^2} \left[
\left(m_i - E_k \right) \left(m_i^2 - m_j^2 + m_k^2 + m_m^2 - 2 m_i
E_k \right) \right. \\
\lefteqn{ \qquad 
\left. \pm p_k \lambda^{\frac{1}{2}} \left((p_i - p_k )^2, m_j^2,
m_m^2 \right)\right] \ ,}
\end{eqnarray}
and $\lambda(x, y, z) = x^2 + y^2 + z^2 - 2(xy + xz + yz)$ as in the
two-body decays.

For our purposes, we analytically reduce this to a one-dimensional
integral over $E_k$ by integrating out $E_m$, and then we evaluate the
integral numerically.  We only evaluate the three-body modes if there
are no kinematically allowed two-body decays\footnote{We always ignore
  two-body decays to gravitino.  See \secref{prog} for details.}.  We
could technically apply the three-body formulas to the entire slepton
sector, but this would in effect double count the on-shell decays of
heavier sleptons.

The full form of these slepton three-body modes is too lengthy to
include here; complete expressions are given in Ref.~\cite{Feng:3body}.
At a qualitative level, there are four distinct decay modes:
\begin{equation}
\tilde{\ell}^- \to \tilde{\ell}^- \ell_k^- \ell_m^+\, , \quad
\nonumber \tilde{\ell}^- \to \tilde{\ell}^+ \ell_k^- \ell_m^-\, , \quad
\nonumber \tilde{\ell}^- \to \tilde{\ell}^+ \nu_k \overline{\nu}_m\, , \quad
\tilde{\ell}^- \to \tilde{\ell}^+ q_k \overline{q}_k \ .
\end{equation}
The first three decays exist in the flavor-conserving case, occurring
through an off-shell neutralino or chargino; the addition of flavor
violating effects introduces flavor-violating couplings and an
interference term between the diagrams in the $\tilde{\ell}^- \to
\tilde{\ell}^+ \ell_k^- \ell_m^-$ mode.  Flavor violating effects also
introduce the possibility of radiating an off-shell $Z$ or Higgs,
which contribute both to the decays to a slepton and a same-generation
lepton and anti-lepton pair and to a slepton and two quarks.


\section{Program Details}
\label{app:program}

Here we outline the structure of the {\tt SPICE} program.  There are
three distinct components: the code to impose generalized
flavor-violating boundary conditions for {\tt SOFTSUSY} and calculate
sleptonic decay widths, an instance of {\tt SUSYHIT} to calculate
non-sleptonic decay widths, and a subroutine which merges the
information from both files into an output file.  As previously
mentioned, the instance of {\tt SUSYHIT} is unchanged from its
original form.

\subsection{{\tt SOFTSUSY} Customization and Sleptonic Decay Widths}

Our customization of the {\tt SOFTSUSY} code has two components ---
renormalization boundary conditions with both gauge and gravity
mediated contributions and calculation of decay widths for slepton
flavor violating decays.

We have not altered the {\tt SOFTSUSY} renormalization algorithms;
however, since the lepton mass matrix is in general non-diagonal, the
mixing between lepton gauge and mass eigenstates must be rotated out
of the lepton sector at the high scale to conform to {\tt SOFTSUSY}'s
boundary conditions.  This is done by first diagonalizing the lepton
mass matrix
\begin{equation}
\ell_{L_\alpha} m_{E_{\alpha \beta}} \ell_{R_\beta} = \ell_{L_\alpha}
U_{\alpha i}^{(\ell_L)} m_{E_{ii}}^D U_{i \beta}^{(\ell_R) \dagger}
\ell_{R_\beta} = \ell_{L_i} m_{E_{ii}}^D \ell_{R_i}\, ,
\end{equation}
where $m_{E_{\alpha \beta}}$ is the original lepton mass matrix and
$m_{E_{ii}}^D$ is the diagonalized lepton mass matrix.  

The interaction terms in the Lagrangian have the form
\begin{equation}
\tilde{\ell}_{L_\alpha}^* \ell_{L_\alpha} = U_{i
\alpha}^{(\tilde{\ell}_L) \dagger} U_{\alpha j}^{(\ell_L)}
\tilde{\ell}_{L_i}^* \ell_{L_j}\, ,
\end{equation}
with similar expressions for the right-handed slepton and sneutrinos.
Then the lepton mixing may be absorbed into the slepton mixing at the
high scale, whereby
\begin{equation}
\begin{array}{c}
U_{i \alpha}^{(\tilde{\ell}_L \dagger)} \to U_{i
\beta}^{(\tilde{\ell}_L) \dagger} U_{\beta \alpha}^{(\ell_L)} \\
\left(M_{\tilde{\ell}_L}^2 \right)_{\alpha \beta} \to   
U_{\alpha \sigma}^{(\tilde{\ell}_L) \dagger}
\left(M_{\tilde{\ell}_L}^2 \right)_{\sigma \rho} U_{\rho
\beta}^{(\tilde{\ell}_L)}\, ,
\end{array}
\end{equation}
again with similar expressions for right sleptons and sneutrinos.

Here we take advantage of the negligible left-right mixing terms at
the high scale to decouple mixing of left and right sleptons, though
the procedure does work in general if the $3 \times 3$ matrix
associated with left-right mixing is also rotated in a similar
fashion.  The diagonalized lepton mass matrix and the rotated slepton
and sneutrino mass matrices are then used as the high-scale boundary
conditions for {\tt SOFTSUSY}'s renormalization routine, and is found
in the {\tt gaugegravityBcs} method.

\subsection{Sleptonic Decay Calculations}

The new calculation of flavor violating decays is encompassed in three
classes: {\tt DecayFormat}, {\tt DecayCalc}, and {\tt
ThreeBodyIntegral}.  Calculations of decay widths is performed in the
{\tt DecayCalc} class, with the help of the {\tt ThreeBodyIntegrals}
class for three-body decays.  The {\tt DecayFormat} class handles
output to the intermediate file and interfaces with the main program.

\subsubsection{{\tt DecayCalc} class}

\begin{center}
\begin{tabular}{ll}
{\bf Method} & {\bf Purpose} \\
\hline
{\tt Lambda} \hspace*{3cm} & Common phase space function \\
\hline
{\tt DecayCalc}           & Loads the slepton masses and mixing matrices \\
                          & from a passed {\tt FlavourMssmSoftsusy} object. \\
\hline
{\tt Alpha, Beta[1-2],}   & Methods which return the value of the \\
{\tt Gamma, Delta[1-2],}  & corresponding coupling coefficient for a given \\
{\tt Sigma[1-4], Rho,}    & set of mass eigenstates.  The methods are in \\
{\tt Zeta[1-2]}           & direct correspondence to the couplings in the \\
                          & Lagrangians presented in \appref{lagrangians}. \\
\hline
{\tt Calc[...]}           & Calculates the decay width (in GeV) of the given \\
                          & decay for a given set of mass eigenstates.  The \\
                          & calculation is performed according to the decay \\
                          & widths given in \appref{twobody} and 
                            \appref{threebody}. \\
\hline
{\tt Output[...]Decays}   & Methods which loop through all decays of a \\
                          & particular particle type and passes decays \\
                          & widths to the DecayFormat object for output.\\
\hline
{\tt IntegralCoefficient} & Methods to produce the beta coefficients \\
{\tt CrossCoefficient}    & which weight the three-body decay integrals.
\end{tabular}
\end{center}

{\tt DecayCalc} loads the slepton mass and mixing matrices from a
passed {\tt FlavourMssmSoftsusy} object, which has already undergone
the renormalization routine.  The slepton and sneutrino masses are
contained in six and three component vectors, respectively, and the
corresponding rotation matrices are $6 \times 6$ and $3 \times 3$
orthogonal matrices.  The rest of the class is dedicated to
calculations of decay widths.  The individual decay widths are
calculated from the methods named {\tt Calc[...]}, which take in the
integer index of the mass eigenstates associated with the decaying and
daughter particles.  Here $i$ is the index of the decaying particle,
while $j$, $k$, and $m$ are the indices of the daughter particles in
order.  Finally the decays are organized in the methods named {\tt
  Output[...]Decays}, which each contain all the decays of the
appropriate parent particle with loops over all decaying and daughter
particle mass indices.  The remainder of the class is composed of
helper functions, including functions corresponding to the functions
Lagrangian coefficients in \appref{lagrangians}.  Additional helper
methods are included in the file {\tt DecayMath.cpp}, which are
primarily functions which perform the integration for three-body
decays.

\subsubsection{{\tt ThreeBodyIntegrals} Class}

\begin{center}
\begin{tabular}{ll}
{\bf Variable} & {\bf Description} \\
\hline
{\tt start}, {\tt end} \hspace*{3cm}
                       & Upper and lower integration limits calculated \\
                       & from the currently saved mass ratios \\
{\tt r[...]}           & Dimensionless ratios of the masses of the \\
                       & daughter particles and the mass of the initial \\
                       & slepton \\
\hline
\hline
{\bf Method} & {\bf Purpose} \\
\hline
{\tt Lambda} \hspace*{3cm} & Common phase space function \\
\hline
{\tt F, F1, F2}            & Common functions used in the integrals \\
\hline
{\tt Simpsons}             & Method to perform an integral of a given \\
                           & one-dimensional function.  The integration \\
                           & limits are stored class variables. \\
\hline
{\tt Integral[1-6]}        & Six integrals corresponding to the standard \\
{\tt CrossIntegral[1-8]}   & three body decay mode and eight corresponding \\
                           & to the intereference term. \\
\hline
{\tt setMasses}            & Sets the mass ratios and integration limits. \\
\hline
{\tt evaluateIntegrals}        & Methods which group together sets of \\
{\tt evaluateCrossIntegrals}   & integrals and then take a sum using a \\
{\tt evaluateCharginoIntegral} & vector of coefficients as weights \\
\end{tabular}
\end{center}

The {\tt ThreeBodyIntegrals} helper class is used to calculate the
values of integrals for the three body decays.  The class takes in
particle masses using the function {\tt setMasses}, which sets common
integration limits and mass ratios to use for multiple integrals.  A
summation over several integrals (or just one for the chargino) can
then be performed using the methods {\tt evaluateIntegrals}, {\tt
evaluateCrossIntegrals}, and {\tt evaluateCharginoIntegral}.
Coefficients for the integrals are passed into these methods to
properly weight the sum.

\subsubsection{{\tt DecayFormat}}

\begin{center}
\begin{tabular}{ll}
{\bf Method} & {\bf Purpose} \\
\hline
{\tt PDGCode} \hspace*{3.5cm} & Function which returns the PDG code of a \\
                              & particle given its type and mass eigenstate \\
\hline
{\tt OutputParticle}      & Output particle code, mass, and lifetime \\
{\tt OutputParticleData}  & Output a table of particle masses and lifetimes \\
\hline
{\tt TwoBodyDecay}        & Output the decay width for a two-body decay \\
{\tt ThreeBodyDecay}      & Output the decay width for a three-body decay \\
{\tt OutputDecayTable}    & Output a table of decay modes with decay widths \\
                          & for decays involving sleptons and sneutrinos \\
\hline
{\tt OutputSusyParams}    & Outputs SUSY parameters to be appended to the \\
                          & end of {\tt HERWIG} input file \\
\hline
{\tt OutputIntermediateFile} & Outputs the full intermediate file containing \\
                             & sleptonic decays
\end{tabular}
\end{center}

Finally, the class {\tt DecayFormat} is devoted to constructing the
intermediate file to send to the {\tt FileCombine} part of the code.
The {\tt DecayFormat} class creates an output stream at declaration
time, and its two methods {\tt TwoBodyDecay} and {\tt ThreeBodyDecay}
are used by the {\tt DecayCalc} class to output two- and three-body
decays, respectively.  The remaining methods are used to output the
particle decay table and {\tt HERWIG} and {\tt SLHA} model parameters.

\subsubsection{Adding a Decay}

The easiest possible alteration to the program is the addition of
another decay mode.  To add a new decay mode, first add a new function
to produce the appropriate decay width in corollary to the functions
{\tt Calc[...]}; this functions should take in the appropriate indices
for the mass eigenstates.  Then place a set of loops in the
appropriate {\tt Output[...]Decays} function, which calculates the
decay width and, if it is non-zero, outputs the decay width through
either the {\tt TwoBodyDecay} or {\tt ThreeBodyDecay} method of the
stored {\tt DecayFormat} object in the {\tt DecayCalc} class.  From
this, the decay mode will be added to the intermediate file and be
automatically included in the final {\tt HERWIG} and {\tt SLHA} output
files.

\subsection{{\tt FileCombine} Program}

The {\tt FileCombine} program merges the intermediate file from the
sleptonic decay calculation and the {\tt SUSYHIT} output.  The program
is composed of three classes: {\tt DecayMode} stores the particle ID's
and decay width of one decay mode.  {\tt Particle} stores all the
information, including a vector of all decay modes, for a single
particle.  {\tt Model} contains a map of all particles, various model
parameters, and both input and output functions.

\subsubsection{{\tt DecayMode} Class}

\begin{center}
\begin{tabular}{ll}
{\bf Method} & {\bf Purpose} \\
\hline
{\tt IsSleptonic} \hspace*{3cm} & Returns true if the parent or any daughter \\
                                & particle is a slepton or sneutrino \\
\hline
{\tt addDaughter}           & Add another daughter particle \\
\hline
{\tt OutputHerwigDecayMode} & Output the decay mode in {\tt HERWIG} format \\
{\tt OutputSLHADecayMode}   & Output the decay mode in SUSY Les Houches 
                              format \\
{\tt OutputSLHA2DecayMode}  & Output the decay mode in SUSY Les Houches 2 
                              format \\
\end{tabular}
\end{center}

The {\tt DecayMode} class stores all information for one decay.  This
includes the parent particle {\tt HERWIG} ID, the number of daughter
particles, the PDG and {\tt HERWIG} ID's of the daughter particles,
and the decay width of the decay mode (in GeV).  The class also
contains methods which output the mode in {\tt HERWIG} and SLHA(2)
format.

\subsubsection{{\tt Particle} Class}

\begin{center}
\begin{tabular}{ll}
{\bf Method} & {\bf Purpose} \\
\hline
{\tt AddDecayMode} & Add given decay mode to the decays for the particle \\
\hline
{\tt FixWidth} & Determine the total decay width based on all currently \\
               & stored decay modes and produce branching fractions \\
\hline
{\tt OutputHerwigDecays} & Outputs the decay table in {\tt HERWIG} format \\
{\tt OutputSLHADecays}   & Outputs the decay table in both SLHA and SLHA2 
                           format \\
\end{tabular}
\end{center}

The {\tt Particle} class contains the mass, total decay width, and
decays for a given particle.  The decays are stored as a vector of
{\tt DecayMode} objects.

\subsubsection{{\tt Model} Class}

\begin{center}
\begin{tabular}{ll}
{\bf Variable} & {\bf Description} \\
\hline
{\tt particleMap} & Map object containing stored particle objects.  The map \\
                  & is indexed by {\tt HERWIG} Id's \\
\hline
\hline
{\bf Method} & {\bf Description} \\
\hline
{\tt PDGToHerwig}        & Converts from PDG to {\tt HERWIG} particle code \\
\hline
{\tt OnShell}            & Checks whether a passed decay mode is on shell \\
\hline
{\tt ReadIntermediate}   & Read in the model and decay information from the \\
                         & intermediate file produced by {\tt SPICE} \\
{\tt ReadLesHouchesFile} & Read in decays from the Les Houches file produced \\
                         & by {\tt SUSYHIT} \\
\hline
{\tt OutputHerwigFile}     & Produces a {\tt HERWIG} input file from the \\
                           & stored model parameters and decays \\
{\tt OutputLesHouchesFile} & Produces a SLHA and SLHA2 output file. \\
                           & This method copies the previous decay file up \\
                           & the decays, then produces a new decay table. \\
\end{tabular}
\end{center}

The {\tt Model} class contains a map of particle objects, stores the
generic model parameters, and handles the input/output functions of
{\tt FileCombine}.  The {\tt ReadIntermediate} function reads in
particle mass information, decays, and model parameters from the
intermediate file; {\tt ReadLesHouchesFile} fills the rest of the
decay table from the {\tt SUSYHIT} output file, discarding off-shell
decays.  The {\tt Combine} function is called from outside the class
to force a consistent order (since the particle mass information is
read from the intermediate file).  The two output functions produce
{\tt HERWIG} input and SLHA and SLHA2 format files.



\end{document}